\PassOptionsToPackage{table}{xcolor}

\documentclass[10pt]{article}
\usepackage[margin=1in]{geometry}
\date{}

\usepackage{pdfpages}
\usepackage{graphbox}
\usepackage{graphicx}
\usepackage{epstopdf}
\usepackage{url}
\usepackage{multirow}
\usepackage{graphicx}
\usepackage{siunitx}
\usepackage{makecell}
\usepackage[table]{xcolor}
\usepackage{cuted}
\usepackage{stfloats}
\usepackage{soul}
\usepackage{xurl}
\usepackage{hyperref}
\usepackage{dcolumn}
\usepackage{bm}
\usepackage{svg}
\usepackage[T1]{fontenc}
\usepackage{authblk}
\usepackage{mathtools}
\usepackage{adjustbox}
\usepackage{placeins}

\providecommand{\keywords}[1]
{\small	
 \textbf{\textit{Keywords---}} #1}

\title{First Steps towards Machine Learning for Prediction and Pre-Correction in Direct Laser Writing}

\author[1,2,3*]{Sven Enns}
\author[1,4]{Julian Hering-Stratemeier}
\author[1,2,3,4,5]{Georg von Freymann}
\affil[1]{Physics Department and Research Center OPTIMAS, RPTU University Kaiserslautern-Landau, Erwin-Schrödinger-Straße 56, 67663 Kaiserslautern, Germany}
\affil[2]{Leibniz University Hannover, Institute of Photonic Systems and Technologies, 30167 Hannover, Germany}
\affil[3]{Leibniz University Hannover, Cluster of Excellence PhoenixD (Photonics, Optics, and Engineering Innovation AcrossDisciplines), Hannover, Germany}
\affil[4]{Opti-Cal GmbH, Erwin-Schrödinger-Straße 56, 67663 Kaiserslautern, Germany}
\affil[5]{Fraunhofer Institute for Industrial Mathematics ITWM, Fraunhofer-Platz 1, 67663 Kaiserslautern, Germany}

\affil[*]{Correspondence to: Sven Enns: \url{sven.enns@qtp.uni-hannover.de}}

\begin{document}
\maketitle
\begin{abstract}
    Additive manufacturing using 2-Photon Polymerization (2PP, aka direct laser writing DLW) enables the fabrication of almost arbitrary complex 3D structures from the meso to the sub-micron scale. 
    However, deviations between the anticipated target structure and the actual print often occur due to physico-chemical processes, limiting the accuracy and reliability of this technology. 
    To minimize these deviations, we hereby present our latest research in developing different neural networks, targeting the above-mentioned aspect. 
    Our networks are trained on several experimental as well as theoretical datasets and show good results in predicting fabrication deviations and (pre-) correcting 2.5D µ-structures. 
    Hence, we demonstrate, that besides conventional iterative correction methods, neural networks are a promising alternative to significantly improving the output quality in DLW. First directly pre-corrected structures show a reduction of the root mean square error up to 93\% in the best case scenario.
    Furthermore, there are no fundamental limitations to transferring this machine learning approach to other 3D printing technologies, as they all face the same challenge in terms of fidelity. 
    To our point of view, the use of neural networks has the potential to enhance the capabilities of this technology, enabling the creation of complex structures with increased accuracy and precision in the near future.
\end{abstract}

\setlength{\tabcolsep}{4pt}

\section{Introduction}\label{sec:intro}
\noindent One of the most high speed and high resolution variants of laser-based additive manufacturing is the 2-Photon Polymerization (2PP) technology, also known as direct laser writing (DLW) first introduced by Maruo \textit{et al.} in 1997 \cite{maruo1997three}. 
There, photo sensitive materials, so-called photo resins, are locally exposed by a (tightly) focused femtosecond laser beam and subsequently cured. 
Depending on the laser power using an objective with high NA ($NA\approx1.4$), feature sizes of less than 100\,nm can be realized \cite{fischer2013three,wollhofen2017functional}. 
On the other hand, the production of macroscopic components within the cm-regime, such as screws and nuts is also possible with this technology \cite{NanoscribeOnline}.
However, a major challenge in this field is the \--- mostly undesired \--- deviation between the envisaged target structure and its corresponding, experimentally 3D printed counterpart. While these deviations are negligible for large structures (e.g. those in the cm-range), they become more significant as the size of the prints decreases.
Unfortunately, this is fundamentally unavoidable with today's photo resins.
Although the principles of DLW have been known for more than 25 years, a holistic calculation of these deviations, based on the physico-chemical processes involved, has been just as impossible as a correct prediction of the complete 3D printed structure. 
Nevertheless, rather simple straight forward calculations of the widths of single lines \cite{fischer2013threeOE} or voxel and pillar dimensions \cite{purtov2019nanopillar} have been published in 2013 and 2019, respectively. 
Taking more physico-chemical processes correctly into account, however, already prevents the prediction of line heights or aspect ratios, as stated in 2022 by Pingali \textit{et al.} \cite{pingali2022reaction}.
A few more approaches focused really successfully on the estimation of printed line dimensions and structure characterization \cite{guney2016estimation,adao2022two}, but, unfortunately still not taking into account all of the most relevant physico-chemical processes.
The complexity and computational intensity of the processes to be modeled still exceed the capacities of today's computers. 
In contrast to such holistic simulation approaches, some research work exists on modeling and investigating partial aspects, such as the molecular diffusion \cite{waller2016spatio}, the so-called Schwarzschild- \cite{yang2019schwarzschild} and proximity-effects \cite{waller2016spatio} or the shrinkage behavior \cite{ovsianikov2009shrinkage}. 
Recently, some of the authors published a new algorithm, covering all the aforementioned physico-chemical processes in a very simplified, approximated way \cite{lang2022towards}. In principle, this approach allows the prediction of arbitrary 2.5D structures, based on seven physical parameters.
Here, the term 2.5D structure refers to those structures that can be represented mathematically by a single 2D matrix $M$ in which each index $M_{ij}$ represents the height of the structure $z_{ij} = z(x,y)$ at the respective lateral coordinate $(x,y)$.
Although the above mentioned prediction approach allows for a pre-correction for the experimentally expected structural deviations in DLW, the approach still depends on calibration structures and clearly has its limitations (see reference \cite{lang2022towards} for details).\\
\noindent Due to these limitations and the general complexity of the physico-chemical processes and their dependence of many external factors, machine learning (ML) approaches could be a solution.
Numerous studies have investigated the application of these approaches in additive manufacturing (AM) in general. Thereby, they are used in multiple stages of the AM workflow, including material design \--- to optimize geometries, topologies, and predict material properties \--- process optimization during fabrication, and post-processing tasks such as defect detection, quality control, and data analysis. A broad spectrum of ML techniques has been implemented for various fabrication processes \cite{MLforAM_1,MLforAM_2}. To further enhance the capabilities of ML, emerging approaches such as physics-informed machine learning (PIML) incorporate physical knowledge about the system into model architectures and loss functions, thereby improving model interpretability and predictive performance, for example as a counter part to resource demanding physical simulations \cite{PIML}. Nevertheless,  ML approaches need to be adapted for all the different applications.
In the context of DLW, only a limited number of studies have explored the use of ML. Similar to other AM techniques, ML can support various stages of the DLW process. For instance, one study applied ML to calibrate data from a DLW-fabricated force sensor during post-processing, amongst other things also compensating for fabrication deviations \cite{DLWForceSensor}. Only very first publications point roughly into the direction of controlling printing design or DLW parameters to generate better printing results:
Lee \textit{et al.} developed an ML model for the automated detection of quality characteristics during DLW printing \cite{lee2020automated}. 
Here, the model was trained to distinguish between 'cured', 'uncured', 'damaged', and 'illuminated' during the fabrication process, allowing for a practical coarse in-time light dosage adjustment.
However, the deviation between anticipated structure and real 3D print can not be minimized by this approach, since accurate dimensions can not be identified.
Moreover, the deforming processes (i.e. shrinkage) are not completed while fabricating.
In a similar manner, Mourka \textit{et al.} presented in a conference talk a ML model to predict the optimal exposure parameters for 2PP, based on scanning electron images \cite{mourka2020machine}.
Unfortunately, no paper followed yet.
To the best of our knowledge, only Yang \textit{et al.} tried to use ML to improve the geometric compliance in DLW \cite{yang2022machine}.
They show an approach to find and correct for systematic deviations in simple structures, such as lenses and lines, by an ML framework.
Although the average errors could be reduced by 50\,--\,80\,\%, the study used data-driven compensation algorithm for the structure correction and lacks of transferability to complex or even arbitrary DLW geometries.\\
\noindent To the best of our knowledge, no further publications have addressed our ultimate goal: 
providing a neural network with a target design that, in turn, generates a pre-compensated version of it, resulting in a 3D print that reproduces the original geometry as accurately as possible.
Therefore, we report on the ongoing improvement in this context and introduce a very first machine learning based approach for direct structure prediction and pre-compensation in DLW.
We developed, trained and tested various neural networks and demonstrate our most promising attempts so far.\\
In this work, we only refer to DLW as polymerization by two- or multi-photon absorption.


\section{Materials \& Methods}\label{sec:m&m}
\noindent This section first describes the printing and measurement steps in detail. 
It then explains how the measurement data is processed, followed by the development of our neural network architectures. 
Finally, we briefly discuss the chosen loss function of the networks, which is a measure for the accuracy of the network and can be used to prioritize certain features during the training while ensuring error convergence of the network training.

\subsection{Fabricating and Measuring}\label{subsec:fab-meas}
\noindent Within this study, all structures shown have been designed as 2D surface matrices $M_{ij}$.
Converting the 2D surface matrices into the standard 3D printer data type `.stl' allows for a common translation into coordinates for the DLW machine:
If not explicitly described otherwise, we have used the software DeScribe from Nanoscribe to discretize our structures into equidistant axial planes with a so-called slicing distance of 0.1\,µm and each plane into lateral lines along the $x$-axis separated by 0.1\,µm -- called hatching distance. 
The thus generated data is interpreted by the associated 3D printer Photonic Professional GT.
Galvanometric mirrors have been used for lateral, and a piezo stage has been used for axial positioning, as well as a constant writing speed of 20\,mm/s. 
The laser power has been kept constant at 45\,mW whereas the beam always illuminates the complete entrance pupil of a 63x objective with a numerical aperture of 1.4 (Carl Zeiss Microscopy Deutschland GmbH). 
IP-S has been used as photo resin. 
The latter, the DLW machine, as well as the slicing software are products of Nanoscribe GmbH \& Co. KG.
The development steps after printing onto ultra-sonic cleaned and subsequently silanized glass substrates \cite{liu20183d} followed the manufacturer's specifications \cite{NanoGuide}: 
first, resting in propylene glycol methyl ether acetate (PGMEA) for 20 minutes, afterwards resting in isopropanol for five minutes, and finally drying gently with nitrogen.\\
For measuring the structure's topography, a µSurf confocal microscope (NanoFocus AG) equipped with a $100\times$ objective ($\text{NA}=0.95$) and a $60\times$ objective ($\text{NA}=0.9$) (Olympus Europa SE \& Co. KG, both) have been used. 
To obtain the best possible results, the microscope's detection parameters, such as exposure time and gain, have been optimized for each single measurement.

\begin{figure}[!htb]
    \centering\includegraphics[width=0.95\textwidth]{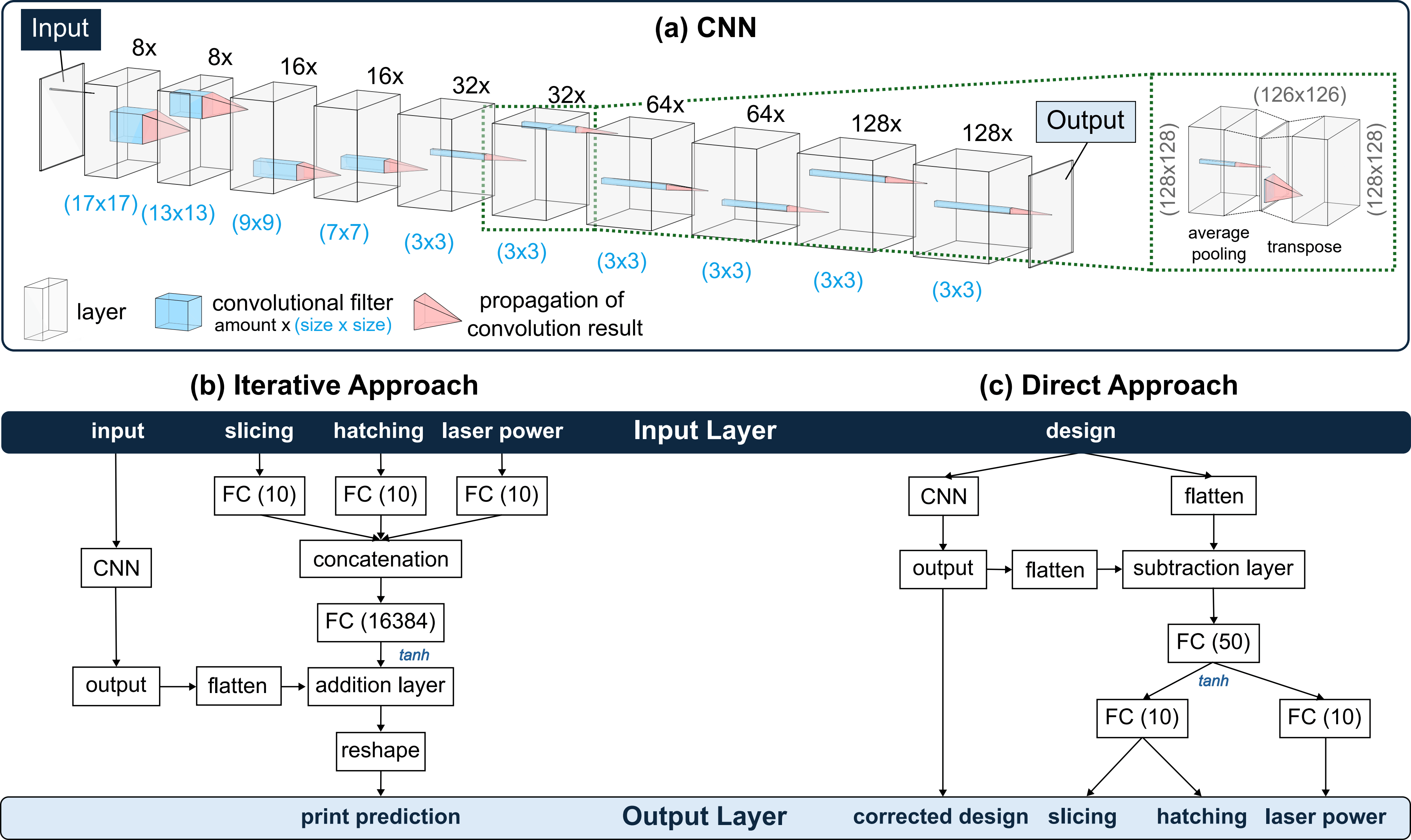}
    \caption{\textbf{Schematic visualization of the network architectures.} 
    (a) General architecture of the neural networks with only convolutional layers.
    Optionally, e.g., pooling layers were implemented as layer substitution by a convolutional, an average pooling, and a deconvolutional layer, as shown in the green inset. 
    (b) \textbf{Iterative Approach} network for structure prediction including the three DLW parameters slicing, hatching, and laser power by a second branch.
    Therefore, also fully connected layers (FC) are used. 
    Since the FC-outcome can be negative, a hyperbolic tangent is used as activation function, whereas all other layers still use rectified linear unit (ReLU) activation. 
    (c) \textbf{Direct Approach} architecture, designed to correct for the printing deviations and predicting a corrected design with corresponding writing parameters. 
    The field 'CNN' represents the network shown in (a).}
    \label{fig:NetworkArchitecture}
\end{figure}

\subsection{Data Preparation}\label{subsec:dataprep}
\noindent The fabrication and measurement techniques described before are used to generate training and test data for our neural networks (see Sec.\,\ref{subsec:Architecture}). 
Since the network training with 3D structures needs much more computational resources and is more time consuming and furthermore a suitable measurement technique to generate accurate training data is not available, different 2.5D structure designs are chosen to build a training data-pool from scratch. 
Besides including, e.g., simple blocks and step structures into the training data-pool, one focus here is on the fabrication of calibration geometries, such as AIR \cite{AIR} or radially symmetric chirped $\text{CIN}_{\text{r}}$ structures \cite{CIN_r} according to ISO 25178–70 \cite{AIR-ARS-ISO}, where a sophisticated structural conformity is crucial, as they have been proven to be fabricable with DLW \cite{Eifler.2018}.
A full list of fabricated structures including exemplary illustrations is shown in the appendix \ref{sec:Training Data}. 
By measuring the height of the final structures, all processes that can lead to deviations are completed. 
Hence, the deviation mechanisms are fully included into the training data of the network. 
Since the fabrication and measurement is time consuming and a large data-pool is needed in order to perform a successful training of the network, additional data is created using data augmentation. 
Data augmentation is an often used technique to increase the amount of training data by creating new data on basis of already existing datasets \cite{Survey-DataAugmentation}. 
In our work, data augmentation can be used by rotating or mirroring existing non-symmetric datasets. 
However, this process is limited not only to the degree of symmetry. 
Since the structure is additively fabricated by printing single lines, a writing direction is introduced which can lead to an anisotropy that has to be taken into account while using data augmentation. 
Hence, data augmentation by rotation is limited to the rotation of 180 degrees.

Starting from the measurement data obtained by confocal microscopy, the following data processing steps are performed, based on the steps by Eifler \textit{et al.} \cite{Eifler.DataProcessing}: 
First, the correct height of the structure has to be determined. 
Therefore, the derivation of the Abbott-Firestone curve \cite{Abbott1933} is calculated. 
Due to the height difference between substrate and structure, it always shows one very prominent jump, allowing for the anticipated height correction.
Further, non-measured points are interpolated, methodically based on the type of structure, e.g., nearest neighbour for step height structures and linear for curved structures. 
Second, tilt and rotation errors due to positioning in the printer and microscope have to be corrected.
Since the base size of the used structures is square, the rotation can be corrected by aligning the edges of the structures.
Correcting the tilt is more complex, as a simple tilt correction of the microscope data based on the measured tilt of the substrate does not take into account the tilt of the substrate during the writing process. 
In order to correct for all tilting errors as precisely as possible, the measurement is compared to the designed structure by calculating a deviation matrix, representing the height difference between the design and the height corrected measurement data. 
For this difference matrix a regression plane can be calculated and used to finally correct the tilt within the measurement data. A schematic illustration of the described steps is shown in the appendix in Figure \ref{fig:DataProcessingScheme}.\\
As normalization is used in neural networks to enhance numerical stability and convergence of the training, the height information is normalized to values between 0 and 1 given the largest structure in the training data. After the training is completed, the result of the network is multiplied by the normalization factor to regain absolute height information. 
The images are scaled to a size of $128\times128$\,pixels. Given the structure base size of $(50\times50)$\,µm$^{2}$ and the confocal microscope's lateral resolution of 0.3125\,µm, this leads to a resolution of 0.391\,µm for further treatment.
Image sizes of $128\times128$\,pixels are chosen as a trade-off to keep the computation time as low as possible without losing too much information by reducing the resolution. 
Thereby an axial resolution of down to 1.5\,nm \cite{NanoFocus} can be achieved. 
The combination of a structure design and its corresponding processed measurement result is a single dataset for our neural networks, where either the design or the measured print can be used as either input or label.
Based on these procedures and tools, a data-pool of in total 5217 labeled datasets is generated.
Different subsets of this data-pool are used depending on the trained network. 
The first subset, used to test the general suitability of the networks, consists of 628 datasets.
Thereby, the size of the structures used to train the network is 50\,µm by 50\,µm with a height between 1\,µm and 25\,µm. 
The base size is kept constant since the image size of our network has to be constant.
A scaling of different structure sizes to the same image size would also scale and, hence, stretch or compress the occurring deviations, leading to a higher inaccuracy of the network.
Later on, more structures and varying printing parameters are introduced to increase the data-pool. 
To get a consistent result, visibly, unusually distorted or defect structures were filtered and not used for the training, while measurement artefacts especially at the edges were kept for this study.
The subsets used for the training are again split into so-called training (85\%) and test data (15\%). 
These are used to check the network's performance and to recognize possible over- or under-fitting. 
To compare the results of different networks and training realizations, a set of structures is defined as benchmark.
In the result section we use blocks, $\text{CIN}_{\text{r}}$- and AIR-type structures, and a logo of the former University of Kaiserslautern to visualize different aspects of the network results.
Thereby, blocks, $\text{CIN}_{\text{r}}$- and AIR-type are structures included into the training but with not trained parameters.
The University's logo is only included into the test data, hence, allows for analyzing the performance of the network on not trained structures.    
To estimate a reachable accuracy of the network prediction, the repetition accuracy of the data generation is tested by multiple executions for different structure types, leading to a deviation of up to 300\,nm.
Since not every structure can be investigated, the accuracy is estimated with a puffer to be about 500\,nm. 
This accuracy can also be regarded as the best possible accuracy that can be expected from the neural network.

\subsection{Network Architecture}\label{subsec:Architecture}
\noindent In the following, two different approaches are presented. 
The first approach aims at the target-print-deviation: 
integrating the design into a neural network, which then predicts the printed outcome. 
Having successfully learned this deviation, the network prediction can then be used to perform a design correction. 
Therefore the following correction mechanism is used analogous to Lang \textit{et al.} \cite{lang2022towards}:
\begin{equation}
	\label{eq:IterationEins}
	M_{k} = M_{\mathrm{Design}} - \alpha \cdot ((M_{\mathrm{Prediction}}-M_{\mathrm{Design}})\ast g_{\mathrm{corr}}),
\end{equation}
\begin{equation}
	\label{eq:IterationZwei}
	M_{k+1} = M_{k} - \alpha \cdot ((M_{\mathrm{Prediction},k}-M_{\mathrm{Design}})\ast g_{\mathrm{corr}}),
\end{equation}
with $M$ as surface matrix, $k$ as iteration numerator, $g_{\mathrm{corr}}$ denoting a two-dimensional Gaussian distribution, and $\alpha$ as correction strength factor. 
Since this step can be performed multiple times we call it the \textbf{Iterative Correction} (\textbf{IC}) approach from now on.
The second approach is called \textbf{Direct Correction} (\textbf{DC}), since the goal is a direct prediction of a corrected design. 
Technically speaking, input and output are switched in comparison to the \textbf{IC}.
In this case, the measured prints of the fabricated training datasets are given as input and the original designs as output.
If now an anticipated design is given as input, the network will predict a corrected version of the design to get a printed structure as close as possible to the input design.\\
In order to analyze image data with neural networks in terms of image classification, object detection and so on, the most popular approach is the application of convolutional neural networks (CNNs) which will also be used in this work. 
We use supervised learning, where every input is labeled. 
In our case, the labeling is done by connecting the datasets of a design with the measurement of its printed result.
Input and label can be switched as explained for the two different approaches.
One of the crucial parts in using neural networks is to find a suitable network architecture for the problem at hand.
To identify a starting point, we use the well-known CNN architectures AlexNet \cite{AlexNet} as well as VGG16 and VGG19 \cite{VGG} as inspiration.
Hence, we present three different network architectures depending on used parameters and tasks.
The first logical task is to predict the target-print-deviation.
Based on this prediction, one can perform the above-mentioned \textbf{Iterative Correction} to get as close as possible to the target print.
The task of the second architecture is the \textbf{Direct Correction} of a corrected design. 
In both cases, only different designs were initially used for training, while the DLW parameters remained constant for the beginning.
Figure\,\ref{fig:NetworkArchitecture}\,(a) shows the basic network architecture for the training without DLW parameters consisting only of convolutional layers.
Decreasing the filter sizes from $17\times17$ to $3\times3$ with a stride of 1 while increasing the layer numbers from 8 to 128, allows to capture larger as well as more detailed features.
As visualized, input and output of the CNN network have to be of same size.
In this case, this is true for every single layer by using zero-padding.
However, this can lead to boundary effects.
Therefore, we add average-pooling layers in between convolutional layers to reduce the strength of these effects.
Since zero padding would lead to a huge average difference on the edges for every pooling layer, no padding is used, leading to a reduction of the output size of this layer.
To regain the initial size, deconvolutional layers, also called transpose layers, are implemented as represented in the inset in Fig.\,\ref{fig:NetworkArchitecture} (a). 
While the network predictions are similar, especially when comparing the loss function values, there are characteristics for each network, as exemplary shown in the appendix\,\ref{sec:PaddingInfluence} for the mentioned boundary effects.
For all layers we choose the rectified linear unit (ReLu) as activation function. 
All parameters, such as filter size or stride, are chosen empirically to find the best possible combination. 
The shown architecture can be used for both, the \textbf{Iterative Correction} and the \textbf{Direct Correction} by switching the input and output.
Since the fabrication parameters have a strong influence on the printing result, the network has to be adjusted in order to capture these effects.
In the following, three crucial parameters -- namely slicing distance, hatching distance and laser power -- are taken into account.
In general, not only the laser power but the exposure dose is a crucial parameter. Thereby, the exposure dose is determined by laser power, pulse duration, repetition rate and scan speed \cite{Skliutas}. While for future models the exposure dose could be one of the network parameters, the present study focuses on the laser power to control the light intensity. Since we want to print as fast as possible for maximum efficiency, the scan speed is kept constant.
\FloatBarrier
\begin{figure}[!ht]
    \centering
    \includegraphics[width=0.82\textwidth]{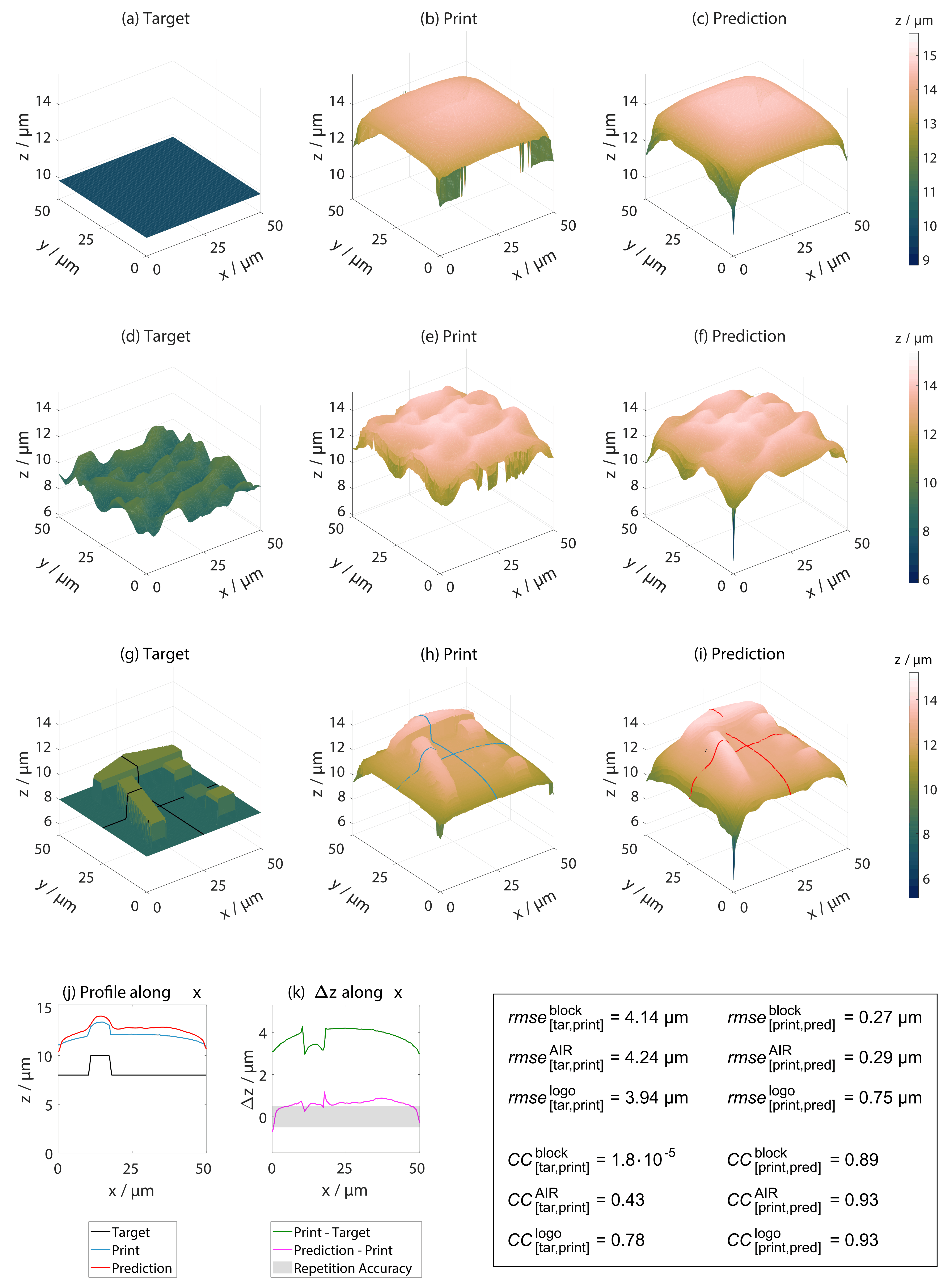}
    \caption{\textbf{Exemplary neural network predictions.}
    The target designs (a, d, g) are compared with the respective measured prints (b, e, h) and the respective network's predictions for a block (a-c), an AIR-type structure (d-f) and the logo of the former University of Kaiserslautern (g-i). 
    For the logo, (j) and (k) show profiles along the $x$-axis and the corresponding height difference $\Delta z$ between print and target, as well as print and prediction. 
    The estimated repetition accuracy of 500\,nm (see Sec.\,\ref{subsec:dataprep}) is shaded in gray. 
    The inset shows a table of the corresponding $rmse$ and correlation coefficients $CC$ to quantify the respective accuracies.
    The neural network was trained with 628 datasets. } 
    \label{fig:Prediction_Examples}
\end{figure}
Thereby, changing the laser power in a certain range leads approximately only to a change of the average structure height, changing hatching and slicing distances on the other hand lead to rather strong and non-symmetric changes. These stem especially from the direction dependent hatching. Further anisotropies can for example be introduced by the polarization of the light \cite{Torok:95,Torok:95_errata,polarization}. However, the effects are shown to be on the scale of nanometers, which is much smaller than the accuracy of the presented model and can hence be neglected at this stage of development. For this reason, circular polarization is chosen for symmetry reason, as is the standard configuration of the commercial system.
The updated network architectures to include the three parameters are shown in Fig.\,\ref{fig:NetworkArchitecture} (b) and (c). 
While for the deviation prediction the parameter information is included in the input layer (b), the \textbf{DC} (c) delivers a corrected design and the corresponding DLW parameters in the output layer.
In both cases, the 'CNN-layers' represent the network architecture from (a). The underlying idea behind these architectures is to create a second branch to process the DLW parameters.
For the prediction, the parameters are converted to a matrix with the size of the input image, representing an parameter dependent height offset, which is then added to the initial CNN prediction.
Since the offset can contain negative values, a hyperbolic tangent is used as activation function for the fully connected layer before the addition step.
In the case of \textbf{DC}, the approach is similar, except that this time the difference between the initial CNN prediction and the target design is calculated, and the appropriate DLW parameters are determined from this.

\begin{table*}[!b]
    \centering
    \caption{\textbf{Quantities of the neural network's performance.} 
    Comparison between target structure, printed structure, the network's prediction of a corrected print, and a corrected print for a block, $\text{CIN}_{\text{r}}$-type, and logo (illustrated in Fig.\,\ref{fig:DirectCorrection}). 
    The comparison is quantified by Pearson correlation coefficient $CC$ (see Eq.\,\ref{eq:Pearson}) and root mean square error $rmse$.}
    \label{tab:DC_Comparison}
    \begin{tabular}{r|cc|cc|cc}
        \hline\hline
        & \multicolumn{2}{c|}{\textbf{block}} 
        & \multicolumn{2}{c|}{\textbf{$\text{CIN}_{\text{r}}$}} 
        & \multicolumn{2}{c}{\textbf{logo}} \\
        & $rmse$ / µm & $CC$
        & $rmse$ / µm & $CC$
        & $rmse$ / µm & $CC$ \\
        \hline
        $[\text{target, print}]$ & 4.14 & 1.8e-05 & 3.67 & 0.27 & 3.49 & 0.78 \\
        $[\text{target, predicted correction}]$ & 0.24 & -4.6e-06 & 0.59 & 0.80 & 0.77 & 0.86 \\
        $[\text{target, printed correction}]$ & 0.27 & -3.1e-05 & 0.58 & 0.88 & 0.90 & 0.87 \\
        \hline\hline
    \end{tabular}
\end{table*}
\subsection{Loss Function}\label{sec:Lossfunction}
\noindent So-called loss functions are defined to evaluate the train and test performances of neural networks and allow for the comparison between different architectures.
Furthermore, choosing a loss function is essential for the convergence of the network results and has an effect on which features are more important along the training process.
As illustrated in Appendix \ref{sec:LossFunction-MSE} using the MSE as an example, not every loss function is suitable for the problem at hand.
Therefore, we introduce a customized loss function (see Eq.\,\ref{eq:Loss}) consisting of the following terms:
\begin{equation}
	L=1-CC+\frac{|\overline{M}_{\mathrm{pred}}-\overline{M}_{\mathrm{lab}}|+\sum_{i,j=1}^n|M^{i,j}_{\mathrm{pred}}-M^{i,j}_{\mathrm{lab}}|}{1.5\cdot\overline{M}_{\mathrm{lab}}}
	\label{eq:Loss}
\end{equation} 
where $CC$ is the Pearson correlation coefficient \cite{pearson1895vii}:
\begin{equation}
	CC=\frac{\sum_{i,j=1}^n(M^{i,j}_{\mathrm{pred}}-\overline{M}_{\mathrm{pred}})\cdot(M^{i,j}_{\mathrm{lab}}-\overline{M}_{\mathrm{lab}})}{\sqrt{\sum_{i,j=1}^n(M^{i,j}_{\mathrm{pred}}-\overline{M}_{\mathrm{pred}})^2\cdot\sum_{i=1}^n(M^{i,j}_{\mathrm{lab}}-\overline{M}_{\mathrm{lab}})^2}}.
	\label{eq:Pearson}
\end{equation}
The correlation coefficient $CC$ can take values between 1 for perfect correlation and -1 for perfect anti-correlation, and is implemented in such a way that even small surface features are given greater importance during the training process.
Since the loss function will be minimized during the training process, we define the first part of the loss function as $L_1=1-CC$ so that $L_1\xrightarrow{}0^+$ in case of perfect correlation between prediction and label, and $L_1\xrightarrow{}2$ for anti-correlation.  
Second, the absolute value of the difference between the mean height value of prediction and label is calculated and added so that the average height of the predicted structure matches the label.
As last term of the loss function every difference between corresponding surface matrix elements of label and prediction are added.
The latter two terms are then divided by the mean height of the label to keep the loss function dimensionless.

Finally, an empirically determined scaling factor of 1.5 is added to the denominator.
As the goal of the network training is the minimization of the prediction error, several optimization methods exist to find the global minimum of the loss function. Thereby, stochastic gradient descent (SGD) methods are very common for CNNs. Several variations of the SGD have been developed in order to decrease convergence time and the probability of getting trapped in a local instead of the global minimum. A test of seven often used methods is shown in \cite{OptimizerTest}. One way to increase convergence speed is the modification of the gradient for each step based on previously calculated gradients. Exemplary, the optimizer Adam (adaptive moment estimation) updates the parameters based on estimated first and second moments, taking into account a moving average of previous gradients with exponentially decaying rates \cite{Adam,KerasAdam}.   
We chose the optimizer algorithm Nadam for the training process, which is similar to Adam but includes the Nesterov moment, describing the calculation of the gradient not at the current parameter set, but one step ahead according to the currently accumulated moment to get a better correction for the next iteration step \cite{Nesterov,Nadam,KerasNadam}.
While the suggested default value for the learning rate, defining the step size for the gradient descent of the optimizer, is 0.001 for Adam and Nadam \cite{Adam, KerasNadam}, we implement a smaller learning rate of 0.0001, as it results in better convergence in our case. Thereby a large step size can lead to overshooting and a small step size increases the computation time.
Resulting training duration and used hardware and programming libraries are also shown in appendix\,\ref{sec:Computation}.

\begin{figure}[!ht]
    \centering
    \includegraphics[width=0.9\textwidth]{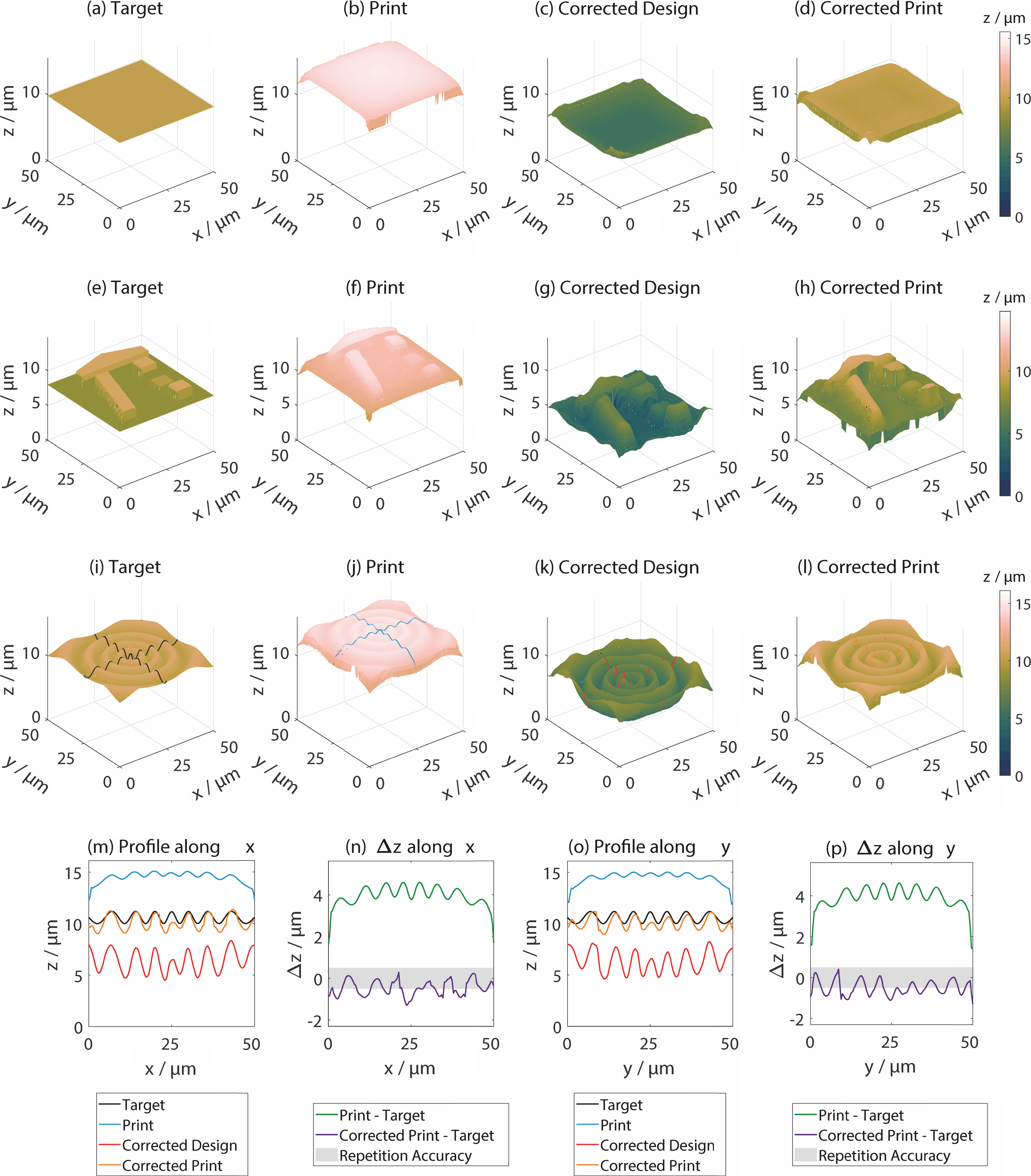}
    \caption{\textbf{Results of the Direct Correction (DC) approach.}
    Examples for the direct design correction of a network trained with 628 datasets. Shown are the corrections of a block (a)-(d), the logo (e)-(h), which was not included in the training data, and a $\text{CIN}_{\text{r}}$-type structure (i)-(l) including profile information (m)-(p).}
    \label{fig:DirectCorrection}
\end{figure}

\section{Results \& Discussion} \label{sec:results}
\noindent In this chapter we present some of the results that we obtained for the defined benchmark structures, giving an overview over achieved accuracy and prominent occurring effects.

\begin{figure}[!ht]
    \centering\includegraphics[width=0.95\textwidth]{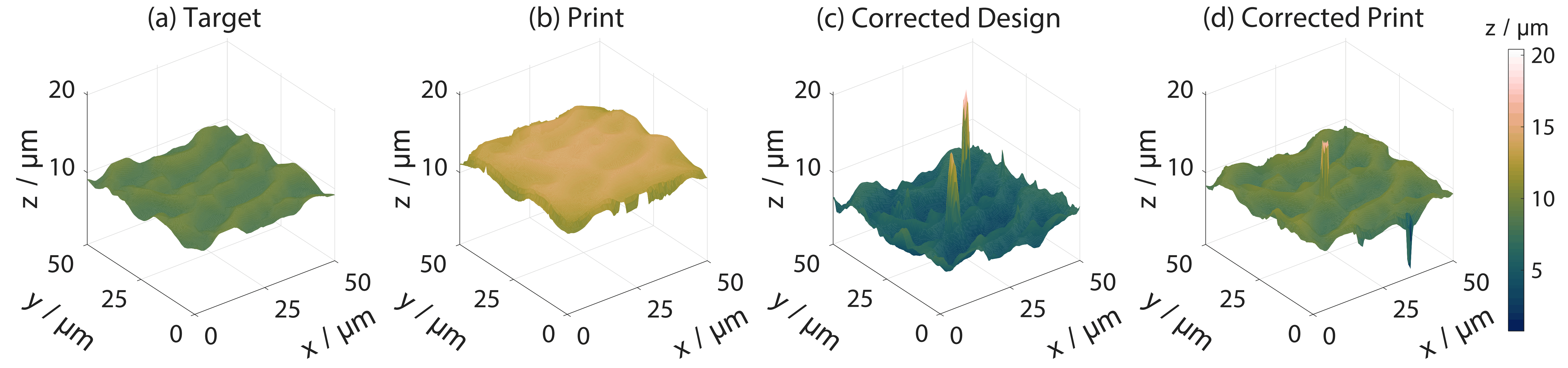}
    \caption{\textbf{Artefacts illustration.}
    Exemplary appearance of a random artefact during the correction of an AIR-type structure (a). 
    While the corrected design (c) of the neural network shows two peaks in the middle of the structure, only one of those is visible in the measurement of the print (d).}
    \label{fig:DirectCorrection_RandomArtefact}
\end{figure}
The first networks (Fig.\,\ref{fig:NetworkArchitecture}) are designed to predict the occurring deviations for different structure types. 
Moreover, they can be used to perform the before mentioned \textbf{Iterative Correction} approach:
three examples using a network, trained with 628 datasets without variation of the DLW parameters are shown in Fig.\,\ref{fig:Prediction_Examples}. 
As first example, the measurement of a printed block is compared to the corresponding network prediction (a-c). 
The height difference between target and print, as well as the bulging in the middle and drop off towards the edges of the structures are predicted well, as the correlation coefficient $CC^{\mathrm{block}}_{[\mathrm{print,pred}]} = 0.89$ and root mean squared error $rmse^{\mathrm{block}}_{[\mathrm{print,pred}]} = 0.27$\,µm (text-inset in Fig.\,\ref{fig:Prediction_Examples}) confirm. 
To evaluate structures, where shrinkage, proximity effect and steep edges are more present, Fig.\,\ref{fig:Prediction_Examples} shows additionally an AIR-type structure (d-f) and the logo of the former University of Kaiserslautern (g-i).
Thereby, the latter structure was not included in the training data. 
The correlation coefficients are consistently above 0.9 and underline the good agreement of the prediction with the actual 3D print.
Since the University's logo is the only of those three structures which is not included in the training data, the highest $rmse$-value of 0.79\,µm can be observed here. 
However, the profile plots in (j-k) show an almost constant offset between print and network's prediction, being only slightly out of the estimated range of repetition accuracy. 
Highest deviations can always be seen in areas of steep edges.\\
While the shown results prove the general suitability of neural networks to predict the printing outcome, a further goal is to pre-correct the target design, hence,  being able to print as close to the target structure as possible. 
Therefore, the aforementioned methods of \textbf{Direct} and \textbf{Iterative Correction (DC}, \textbf{IC}, respectively) are applied. 
The \textbf{DC}-approach is the quicker and simpler method, since the input of the target design directly results in a corrected design.
Again, the network's performance is tested by three benchmark structures: a block, a circular chirped $\text{CIN}_{\text{r}}$-type structure -- which are both part of the training data but with untrained parameters -- and the logo. 
The results are visualized in Fig.\,\ref{fig:DirectCorrection}. 
While the improvement is already clearly visible, Tab.\,\ref{tab:DC_Comparison} quantifies once again the $CC$ and $rmse$ values. 
Thereby, the corrected print of the block reaches an $rmse$ similar to the prediction accuracy of the network, shown before. 
For the $\text{CIN}_{\text{r}}$-type structure, especially the increase of the correlation coefficient from 0.27 between target and uncorrected print to 0.88 between target and corrected print, as well as the $rmse$-reduction by about 84\,\% is remarkable. 
As the profile plots in Fig.\,\ref{fig:DirectCorrection} (m-p) show, the difference between target and corrected print is partly in the range of repetition accuracy.
However, the frequency and especially the amplitude of the $\text{CIN}_{\text{r}}$ prints' surface features do still not match the target design sufficiently. 
Furthermore, the correction does not necessarily preserve the symmetry of the structure.
\begin{figure}[b!]
    \centering
    \includegraphics[width=0.9\textwidth]{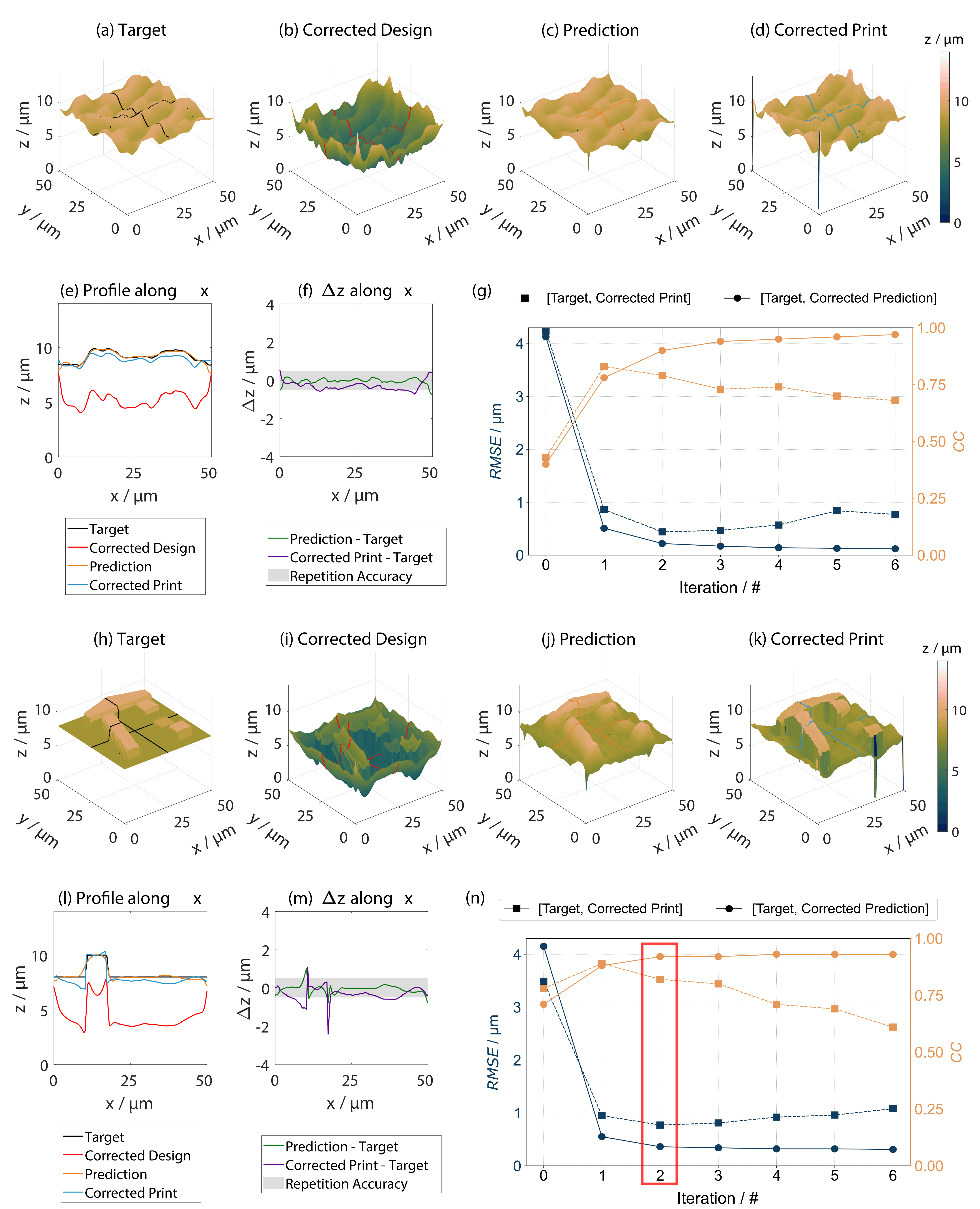}
    \caption{\textbf{Iterative Correction (IC).}
    The best result of an AIR-type structure (a-f) is reached after two iteration steps during the \textbf{IC}-approach. 
    In (g), the $rmse$ and $CC$ values are plotted against the iterations.
    Similarly, (h)-(n) show the results for the logo structure. 
    The uncorrected result is denoted as iteration 0, respectively.
    The neural network was trained with 628 datasets.}
    \label{fig:IterativeCorrection}
\end{figure}
In particular, regarding the pre-correction of such calibration structures, this is a disadvantage in comparison to the algorithmic approach presented by Lang \textit{et al.} \cite{lang2022towards}.
However, also for the non-trained logo, the correction reduces the initial $rmse$ value from 3.67\,µm to 0.9\,µm while increasing the correlation from 0.78 to 0.87. 
As visible in the surface plots, the main deviations are caused by artefacts at the edges and the less steep flanks in comparison to the target. 
Furthermore, the flat surface areas of the target design stay bulged in the corrected print. 
The combination of bulging and edge artefacts can lead to locally strong deformations. 
While an overall improvement is clearly visible and quantitatively confirmed, the feature deformation still can cause problems, depending on the application of the printed structure.\\
During several tests and training processes, the appearance of random artefacts was noticeable, although not quantifiable. 
One example is shown in Fig.\,\ref{fig:DirectCorrection_RandomArtefact}, where a random peak in the middle of the structure appeared.
While the rest of the corrected print shows good agreement with the target, this random artefact leads to increased error values and can render the structure unusable.\\
In contrast to the \textbf{Direct Correction}, the \textbf{Iterative Correction}, using the deviation prediction according to Eq.\,\ref{eq:IterationEins} and \ref{eq:IterationZwei}, is illustrated in Fig.\,\ref{fig:IterativeCorrection}. 
Again, based on the examples of an AIR-type structure (a-g) and the logo structure (h-n). 
\begin{table*}[!t]
    \centering
    \caption{\textbf{Quantitative comparison of Iterative (IC) and Direct Correction approach (DC).}
    Both correction approaches are quantified by mean height difference $\Delta\overline{M}$, height corrected root mean square error $\tilde{rmse}$, Pearson correlation coefficient $CC$ (of the inner 80\,\% of the surface $CC_{80}$), arithmetic and squared surface roughness parameters $S_{\text{a}}$ and $S_{\text{q}}$ and axial amplification coefficient $\alpha_{z}$ (of the inner 80\,\% of the surface $\alpha_{z,80}$). 
    The respective better value is written in green.}
    \label{tab:overall_Comparison}
    \begin{tabular}{r|cccc|cccc}
    \hline\hline
        & \multicolumn{4}{c|}{\textbf{AIR}} 
        & \multicolumn{4}{c}{\textbf{$\text{CIN}_{\text{r}}$}} \\
        & target & print & DC & IC
        & target & print & DC & IC \\
        \hline
        $\Delta\overline{M}$ / µm & 0 & 4.198 & -0.775 & \textcolor{green!60!black}{-0.274} & 0 & 3.614 & -0.389 & \textcolor{green!60!black}{-0.367} \\
        $\tilde{rmse}$ / µm & 0 & 0.605 & 0.380 & \textcolor{green!60!black}{0.344} &  0 & 0.653 & 0.433 & \textcolor{green!60!black}{0.321} \\
        $CC$ & 1 & 0.432 & \textcolor{green!60!black}{0.854} & 0.793 & 1 & 0.266 & \textcolor{green!60!black}{0.878} & 0.812 \\
        $CC_{\text{80}}$ & 1 & 0.774 & \textcolor{green!60!black}{0.935} & 0.911 & 1   & 0.666 & \textcolor{green!60!black}{0.911} & 0.887 \\
        $S_{\text{a}}$ / µm & 0.411 & 0.472 & 0.580 & \textcolor{green!60!black}{0.440} & 0.374 & 0.461 & 0.660 & \textcolor{green!60!black}{0.381} \\
        $S_{\text{q}}$ / µm & 0.474 & 0.633 & 0.694 & \textcolor{green!60!black}{0.564} & 0.418 & 0.625 & 0.751 & \textcolor{green!60!black}{0.490} \\
        $\alpha_{z}$ & 1 & 1,229 & 1.439 & \textcolor{green!60!black}{1.124} & 1 & 1.288 & 1.779 & \textcolor{green!60!black}{1.050} \\
        $\alpha_{z\text{,80}}$ & 1 & 0.735 & 1.376 & \textcolor{green!60!black}{0.987} & 1 & 0.672 & 1.683 & \textcolor{green!60!black}{0.931} \\
        \hline\hline
    \end{tabular}
\end{table*}
\begin{table}[!b]
    \centering
    \caption{\textbf{Quantities of the neural network's performance with DLW parameters.} 
    Comparison of target structure with measurement and network prediction of the corrected print for a block, $\text{CIN}_{\text{r}}$ and logo shown in Fig. \ref{fig:Multi-Input_Prediction}. The network was trained with 5217 datasets including parameter variations.}
    \label{tab:Multi-Input_Prediction}
    \begin{tabular}{r|ccc|ccc|ccc}
        \hline\hline
        & \multicolumn{3}{c|}{\textbf{block}} 
        & \multicolumn{3}{c|}{\textbf{AIR}} 
        & \multicolumn{3}{c}{\textbf{logo}} \\
        & $\Delta\overline{M}$ / µm & $rmse$ / µm & $CC$
        & $\Delta\overline{M}$ / µm & $rmse$ / µm & $CC$
        & $\Delta\overline{M}$ / µm & $rmse$ / µm & $CC$ \\
        \hline
        $[\text{target, print}]$ & 3.22 & 0.79 & 1.6e-5 & 1.86 & 0.49 & 0.61 & 2.59 & 0.49 & 0.84 \\
        $[\text{target, predicted}]$ & 1.06 & 0.27 & 0.94 & 0.56 & 0.18 & 0.96 & 0.65 & 0.25 & 0.96 \\
        \hline\hline
    \end{tabular}
\end{table}
Since for this \textbf{IC}-approach only the structure's design and the DLW parameters are needed, all correction iterations can be prepared at once and printed within one single job, saving time and ensure same printing conditions throughout all iteration steps for good comparability. 
As shown in (g) for the AIR-type structure and in (n) for the logo structure, the network's prediction of the corrected designs show convergent behavior towards higher correlation and lower $rmse$ values for ongoing iteration steps, just as expected. 
However, the measurement data show a different behavior. 
While in both cases, a minimum of the $rmse$ is obtained at the second iteration step, the correlation coefficient undergoes a steady decrease. 
This effect and the again increasing $rmse$ after the second iteration are both results of artefacts. These artefacts get more dominant for every iteration as visualized in appendix\,\ref{sec:Iterative Correction}, showing the corrected prints of the AIR-type structure for the first six iteration steps. 
The artefacts mainly occur at the edges of the structure and at steeper flanks, as has already been identified as problematic for the \textbf{DC}-approach, before. 
Since the artefacts cannot be properly mapped by the network, they can be enhanced by every iteration step. 
While in the shown cases the best corrected print is achieved for the second iteration, this does not have to be the case for every structure. 
Hence, using the \textbf{IC}-approach still demands for the printing of several corrected structures to find the best possible result. 
Anyhow, since these iterations do not rely on the measurements of each other, they can be printed all in one single job, making this approach very efficient.
Fig.\,\ref{fig:IterativeCorrection} presents the second iteration for both structures. 
A good overall agreement between the target and the corrected print is obvious with only few artefacts. 
As the profile plots reveal, the difference lies almost completely in the repetition accuracy range. 
A comparison between the uncorrected print, the \textbf{DC}-approach, and the best result of the \textbf{IC}-approach for the $\text{CIN}_{\text{r}}$-type and AIR-type structures is shown in Tab.\,\ref{tab:overall_Comparison}. 
Here, two different error characteristics are evaluated. 
First, the average height difference relative to the design structure is calculated.
Second, to further analyse the quality of the surface features additionally to the correlation coefficient, the mean height is subtracted for each structure, leaving the surface features.
The $rmse$ is then calculated accordingly. 
Finally, metrological parameters, such as the axial amplification coefficient $\alpha_{\text{z}}$, and the arithmetic and quadratic surface roughness $S_{\text{a}}$ and $S_{\text{q}}$ are compared. 
Thereby the index '80' refers to the percentage of the evaluated surface: only the inner 80\,\% of the surface was evaluated to get rid of edge effects.
As the data show, the \textbf{Iterative Correction} method delivers the best results regarding the design. 
Only for the correlation coefficient, the \textbf{DC}-approach shows slightly higher values. 
However, as visualized in Fig.\,\ref{fig:IterativeCorrection} and appendix\,\ref{sec:Iterative Correction}, the increasing influence of artefacts in the \textbf{IC}-approach is assumed to cause this lower correlation. 
This assumption can be supported by comparing the correlation coefficient over the complete structure $CC$ and the reduced one $CC_{\text{80}}$. 
Although the \textbf{DC}-approach still shows a higher correlation value, the effect of the edge region is stronger for the \textbf{IC}-approach. 
Since the increasing artefacts appear mostly on edges, this can also be the case in steeper regions in the middle of the structure, additionally lowering the correlation.
\begin{figure}[!ht]
    \centering
    \includegraphics[width=0.85\textwidth]{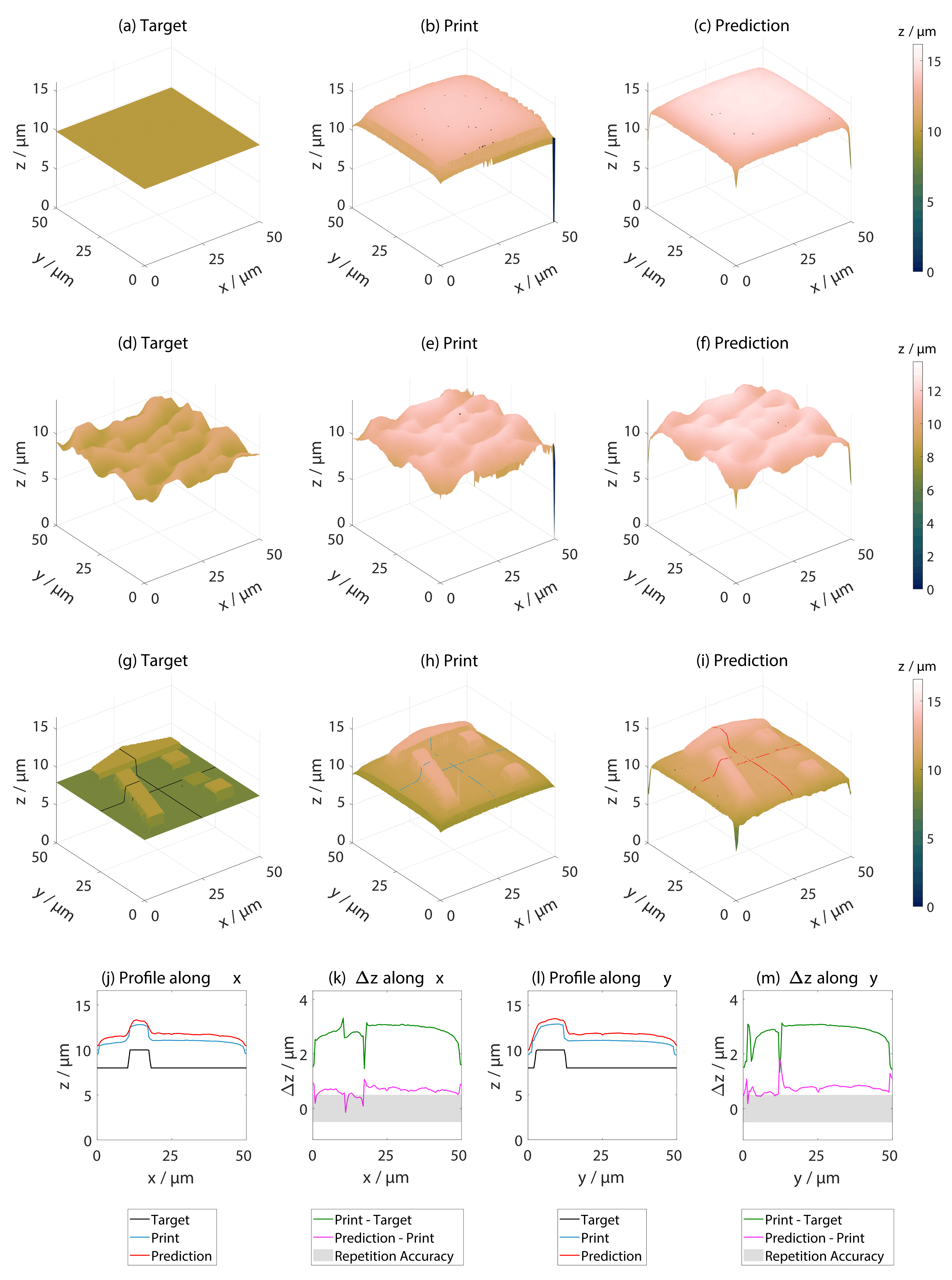}
    \caption{\textbf{Neural network prediction with DLW parameters.}
    Exemplary predictions shown for a block structure (a-c), an AIR-type structure (d-f), and the logo structure (g-m).
    The network was trained with 5217 training datasets.}
    \label{fig:Multi-Input_Prediction}
\end{figure}
\begin{figure}[!ht]
    \centering
    \includegraphics[width=0.85\textwidth]{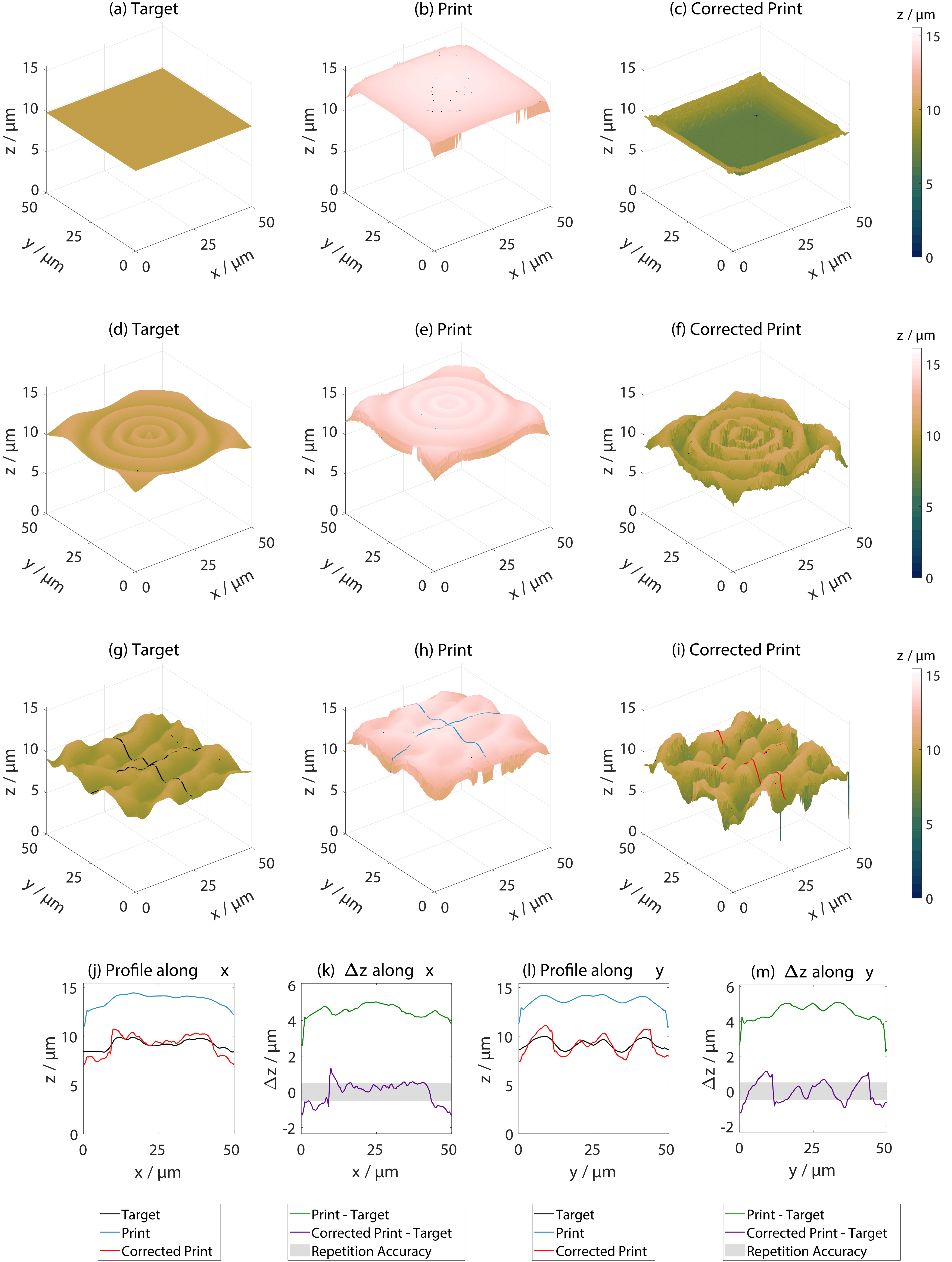}
    \caption{\textbf{Neural network design correction with DLW parameters.}
    Exemplary design corrected printing results, shown for a block structure (a-c), a $\text{CIN}_{\text{r}}$-type structure (d-f), and an AIR-type structure (g-m) in comparison to the non-corrected print.
    The network was trained with 5217 training datasets.}
    \label{fig:DirectCorrection_inclDLWParameter}
\end{figure}

Noteworthy is however, that despite of the lower correlation, the axial amplification coefficient $\alpha_{\text{z}}$ for the \textbf{IC}-approach is closer to the ideal value of 1 and shows additionally the lowest surface roughnesses $S_{\text{a}}$ and $S_{\text{q}}$, even lower than the non-corrected print. 
These results show, that the \textbf{IC}-approach leads to the best possible results, but needs more effort than the \textbf{DC}-approach, since the real error of the corrected structure does not converge with higher iteration numbers.\\ 
\begin{table*}[!t]
    \centering
    \caption{\textbf{Quantitative comparison of Iterative (IC) and Direct Correction approach (DC), with DLW parameters.}
    Analysis of printing results shown in Fig.\,\ref{fig:DirectCorrection_inclDLWParameter}.
    Both correction approaches are quantified by mean height difference $\Delta\overline{M}$, height corrected root mean square error $\tilde{rmse}$, Pearson correlation coefficient $CC$ (of the inner 80\,\% of the surface $CC_{80}$), arithmetic and squared surface roughness parameters $S_{\text{a}}$ and $S_{\text{q}}$ and axial amplification coefficient $\alpha_{z}$ (of the inner 80\,\% of the surface $\alpha_{z,80}$).
    The red and green cells indicate the additional comparison to the results of the \textbf{DC}-approach in Tab.\,\ref{tab:overall_Comparison}, where no DLW parameters have been included.
    Green shows a better and red a worse result, respectively.}
    \label{tab:DirectCorrection_inclDLWParameter}
    \begin{tabular}{r|ccc|ccc}
    \hline\hline
        & \multicolumn{3}{c|}{\textbf{AIR}} 
        & \multicolumn{3}{c}{\textbf{$\text{CIN}_{\text{r}}$}} \\
        & target & print & DC
        & target & print & DC \\
        \hline
        $\Delta\overline{M}$ / µm & 0 & 4.198 & \textcolor{green!60!black}{-0.096} & 0 & 3.614 & \textcolor{red!60!black}{1.065} \\
        $\tilde{rmse}$ / µm & 0 & 0.605 & \textcolor{red!60!black}{0.620} &  0 & 0.653 & \textcolor{red!60!black}{0.691} \\
        $CC$ & 1 & 0.432 & \textcolor{green!60!black}{0.935} & 1 & 0.266 & \textcolor{green!60!black}{0.886} \\
        $CC_{\text{80}}$ & 1 & 0.774 & \textcolor{green!60!black}{0.947} & 1 & 0.666 & \textcolor{red!60!black}{0.893} \\
        $S_{\text{a}}$ / µm & 0.411 & 0.472 & \textcolor{red!60!black}{0.902} & 0.374 & 0.461 & \textcolor{red!60!black}{0.919} \\
        $S_{\text{q}}$ / µm & 0.474 & 0.633 & \textcolor{red!60!black}{1.041} & 0.418 & 0.625 & \textcolor{red!60!black}{1.033} \\
        $\alpha_{z}$ & 1 & 1,229 & \textcolor{red!60!black}{2.179} & 1 & 1.288 & \textcolor{red!60!black}{2.444} \\
        $\alpha_{z\text{,80}}$ & 1 & 0.735 & \textcolor{red!60!black}{2.175} & 1 & 0.672 & \textcolor{red!60!black}{2.540} \\
        \hline\hline
    \end{tabular}
\end{table*}
Since the general functionality and suitability of the neural network has been shown, it now has to be investigated, if the network can also perform well in case of added DLW parameters. 
The updated network architecture to include these parameters into the learning and prediction process is shown in Fig.\,\ref{fig:NetworkArchitecture}\,(b) and (c). 
The respective prediction results are illustrated in Fig.\,\ref{fig:Multi-Input_Prediction} and Tab.\,\ref{tab:Multi-Input_Prediction}. 
While the correlation coefficient between the prints and their associated network predictions show values above 0.94 and are similar to the results without DLW parameters, the $rmse$ tend to be higher. 
Hence, the general ability of the network to predict the exemplary shown structures is still assured, however, larger deviations can be expected.
Thereby, these deviations depend also on the chosen parameter set.
Understandably, this can be attributed to the general amount of training datasets and also their distribution among the different parameter values. 
While the predictions of the network for different DLW parameters show only few artefacts with the largest deviation caused by a difference of the average height, the corrections by the network introduce further effects.
This is noticeable by comparing the mean height differences and height corrected $rmse$s of both approaches in Tab.\,\ref{tab:Multi-Input_Prediction} and Tab.\,\ref{tab:DirectCorrection_inclDLWParameter}. 
While the mean height difference varies in a range of about 1\,µm, the height corrected $rmse$ between the corrected print and the target is about twice as large as between the print and the prediction. 
As shown in Fig.\,\ref{fig:DirectCorrection_inclDLWParameter}, even for a simple block structure the deviation from the design is much larger than for all previously shown approaches.
For all corrected prints, both, the image illustration and the roughness values show an increased roughness of the structure surface.
Moreover, the profile plots not only show a good agreement of the average structure height of the corrected print, but also an overshooting correction for the features.
This results in a stronger variation, observable in the difference plots (k) and (m). 
In contrast to the results before, the edges do not show a worsening effect as correlation coefficients and amplification coefficients stay similar when the edges are not included in the calculation. 
Some of the increased errors, such as the higher roughness of the structures, can be partially caused by the choice of printing parameters, e.g. a too large hatching distance. 
While the prediction approach for a design and a corresponding DLW parameter set should only lead to a prediction inside of a repetition accuracy range, the \textbf{Direct Correction} approach has a theoretically infinite amount of possible design-parameter-combinations.
This results in a higher chance of including error sources which could not be mapped during the training process of the network. 
However, while the red marked values in Tab.\,\ref{tab:DirectCorrection_inclDLWParameter} show worse results in comparison to the \textbf{DC}-results without DLW parameters (Tab.\,\ref{tab:overall_Comparison}), especially for the correlation coefficient a few improved results are noticeable.
This is actually surprising, since the amount of training datasets per parameter-set is lower, hence, underlining the potential of the chosen approach.
A better prediction implies a better suitability of the \textbf{Iterative Correction}-approach in the future, as seen in the studies without including the DLW parameter.
This requires the development of a suitable algorithm to update both, design and parameters at the same time without overshooting.

\section{Summary \& Outlook} \label{sec:outlook}
\noindent In this work, different neural networks were developed to predict and correct for deviations between the anticipated structures and their respective 2PP printed counterparts. 
The networks were trained on both, experimental and theoretical datasets and show generally convincing results for 2.5D structures. 
Our approaches demonstrate that using neural networks is a promising alternative to conventional iterative correction methods. Once the network is trained, directly pre-corrected designs can be printed. Therefore, time consuming iterations can be reduced saving experimental effort. Thereby, the studies of exemplary benchmark structures illustrate a reduction of the root mean square error between 74\% to 93\% in the case without varied DLW parameters. The approach including DLW parameters show reduction of the $rmse$ between 46\% and 85\%, emphasizing the importance of increasing training data sets. Across both cases an increase of the correlation coefficient to the range between 0.79 and 0.93 is achieved. Depending on many factors such as the amount of training data, used structure types and parameters, the result for every structure type might differ strongly. However, the numbers show the potential of this approach, making further development worthwhile.\\
\noindent Since the performance of the network prediction is crucially influenced by the amount of training data, the study is currently restricted to a single device and resin. Thereby, the utilization of different laser sources and devices can yield divergent printing results \cite{Cantoni,Ladika} and resins show characteristic behavior regarding feature sizes, surface roughness and shrinkage. Nonetheless, there is no general physical limitation to extending this approach to a larger number of materials and devices. However, for each added parameter, the amount of necessary training data increases significantly, requiring new studies for proper evaluation and further step by step development of the model. For future work, more investigations are necessary to further improve the performance of neural networks and to extend its applicability to various structure types and sizes, such as real 3D and stitched structures. Additionally, more fabrication parameters, like (varying) hatching directions and polarization, fabrication speed or the numerical aperture of the objective lens are to be implemented to cover a significantly broadened application area in the field of DLW. Furthermore, calibration structures for transferring the neural network outputs to different DLW machines must be developed, to make this machine learning approach beneficial for all DLW working groups.

\section*{Acknowledgements}
    \noindent The authors gratefully acknowledge the institute for measurement and sensor technology (MTS) at the RPTU University Kaiserslautern-Landau for the opportunity of taking confocal measurements and the measuring know how.\\
    Funded by the Deutsche Forschungsgemeinschaft (DFG, German Research Foundation) – Project-ID 172116086 – SFB 926.

\bibliographystyle{ieeetr}
\bibliography{_References.bib}

@article{maruo1997three,
  title={Three\nobreakdash-dimensional microfabrication with two\nobreakdash-photon\nobreakdash-absorbed photopolymerization},
  author={Maruo, Shoji and Nakamura, Osamu and Kawata, Satoshi},
  journal={Optics letters},
  volume={22},
  number={2},
  pages={132--134},
  year={1997},
  publisher={Optica Publishing Group},
  doi={10.1364/ol.22.000132}
}

@article{fischer2013three,
  title={Three\nobreakdash-dimensional optical laser lithography beyond the diffraction limit},
  author={Fischer, Joachim and Wegener, Martin},
  journal={Laser \& Photonics Reviews},
  volume={7},
  number={1},
  pages={22--44},
  year={2013},
  publisher={Wiley Online Library},
  doi={10.1002/lpor.201100046}
}

@article{wollhofen2017functional,
  title={Functional photoresists for sub\nobreakdash-diffraction stimulated emission depletion lithography},
  author={Wollhofen, Richard and Buchegger, Bianca and Eder, Christine and Jacak, Jaroslaw and Kreutzer, Johannes and Klar, Thomas A},
  journal={Optical Materials Express},
  volume={7},
  number={7},
  pages={2538--2559},
  year={2017},
  publisher={Optica Publishing Group},
  doi={10.1364/OME.7.002538}
}

@misc{NanoscribeOnline,
  author = 	 {Rushabh Haria},
  title = 	 {{Nanoscribe Photonic Professional GT 3D printer produces microscopic molds in series.}},
  year = 	 2017,
  note = 	 {at \url{https://3dprintingindustry.com/news/3d-printer-nanoscribe-photonic-polymer-molds-125041/}}
}

@article{fischer2013threeOE,
  title={Three\nobreakdash-dimensional multi\nobreakdash-photon direct laser writing with variable repetition rate},
  author={Fischer, Joachim and Mueller, Jonathan B and Kaschke, Johannes and Wolf, Thomas JA and Unterreiner, Andreas-Neil and Wegener, Martin},
  journal={Optics Express},
  volume={21},
  number={22},
  pages={26244--26260},
  year={2013},
  publisher={Optica Publishing Group},
  doi={10.1364/OE.21.026244}
}

@article{purtov2019nanopillar,
  title={Nanopillar diffraction gratings by two\nobreakdash-photon lithography},
  author={Purtov, Julia and Rogin, Peter and Verch, Andreas and Johansen, Villads Egede and Hensel, Ren{\'e}},
  journal={Nanomaterials},
  volume={9},
  number={10},
  pages={1495},
  year={2019},
  publisher={Multidisciplinary Digital Publishing Institute}
}

@article{pingali2022reaction,
  title={Reaction\nobreakdash-Diffusion Modeling of Photopolymerization During Femtosecond Projection Two\nobreakdash-Photon Lithography},
  author={Pingali, Rushil and Saha, Sourabh K},
  journal={Journal of Manufacturing Science and Engineering},
  volume={144},
  number={2},
  pages={021011},
  year={2022},
  publisher={American Society of Mechanical Engineers Digital Collection}
}

@article{guney2016estimation,
  title={Estimation of line dimensions in 3D direct laser writing lithography},
  author={Guney, M. G. and Fedder, G. K.},
  journal={Journal of Micromechanics and Microengineering},
  volume={26},
  number={10},
  pages={105011},
  year={2016},
  publisher={IOP Publishing}
}

@article{adao2022two,
  title={Two\nobreakdash-photon \text{polymerization} \text{simulation} and fabrication of 3D \text{microprinted} \text{suspended} waveguides for on\nobreakdash-chip optical \text{interconnects}},
  author={Ad{\~a}o, Ricardo M. R. and Alves, Tiago L and Maibohm, Christian and Romeira, Bruno and Nieder, Jana B},
  journal={Optics Express},
  volume={30},
  number={6},
  pages={9623--9642},
  year={2022},
  publisher={Optica Publishing Group}
}

@article{waller2016spatio,
  title={Spatio\nobreakdash-temporal proximity characteristics in 3D µ\nobreakdash-printing via multi\nobreakdash-photon absorption},
  author={Waller, Erik Hagen and von Freymann, Georg},
  journal={Polymers},
  volume={8},
  number={8},
  pages={297},
  year={2016},
  publisher={MDPI},
  doi={10.3390/polym8080297}
}

@article{yang2019schwarzschild,
  title={On the schwarzschild effect in 3D two\nobreakdash-photon laser lithography},
  author={Yang, Liang and M{\"u}nchinger, Alexander and Kadic, Muamer and Hahn, Vincent and Mayer, Frederik and Blasco, Eva and Barner-Kowollik, Christopher and Wegener, Martin},
  journal={Advanced Optical Materials},
  volume={7},
  number={22},
  pages={1901040},
  year={2019},
  publisher={Wiley Online Library},
  doi={10.1002/adom.201901040}
}

@article{ovsianikov2009shrinkage,
  title={Shrinkage of microstructures produced by two\nobreakdash-photon polymerization of Zr\nobreakdash-based hybrid photosensitive materials},
  author={Ovsianikov, Aleksandr and Shizhou, Xiao and Farsari, Maria and Vamvakaki, Maria and Fotakis, Costas and Chichkov, Boris N},
  journal={Optics Express},
  volume={17},
  number={4},
  pages={2143--2148},
  year={2009},
  publisher={Optica Publishing Group},
  doi={10.1364/OE.17.002143}
}

@article{lang2022towards,
  author = {Nicolas Lang and Sven Enns and Julian Hering and Georg von Freymann},
  journal = {Optics Express},
  number = {16},
  pages = {45302},
  publisher = {Optica Publishing Group},
  title = {Towards efficient structure prediction and pre\nobreakdash-compensation in multi\nobreakdash-photon lithography},
  volume = {30},
  year = {2022},
  doi={10.1364/OE.462775}
}

@article{MLforAM_1,
  title={Machine learning in additive manufacturing: State\nobreakdash-of\nobreakdash-the\nobreakdash-art and perspectives},
  author={C. Wang and X.P. Tan and S.B. Tor and C.S. Lim},
  journal={Additive Manufacturing},
  volume={36},
  pages={101538},
  year={2020},
}

@article{MLforAM_2,
  title={Research and application of machine learning for additive manufacturing},
  author={Qin, Jian and Hu, Fu and Liu, Ying and Witherell, Paul and Wang, Charlie C.L. and Rosen, David W. and Simpson, Timothy W. and Lu, Yan and Tang, Qian},
  journal={Additive Manufacturing},
  volume={52},
  pages={102691},
  year={2022},
}

@article{PIML,
  title={A review on physics\nobreakdash-informed machine learning for process\nobreakdash-structure\nobreakdash-property modeling in additive manufacturing},
  author={Faegh, Meysam and Ghungrad, Suyog and Oliveira, João Pedro and Rao, Prahalada and Haghighi, Azadeh},
  journal={Journal of Manufacturing Processes},
  volume={133},
  pages={524-555},
  year={2025},
}

@article{DLWForceSensor,
  title={Micro\nobreakdash-scale fiber\nobreakdash-optic force sensor fabricated using direct laser writing and calibrated using machine learning},
  author={Thompson, Alex J. and Power, Maura and Yang, Guang-Zhong},
  journal={Optics Express},
  volume={26},
  pages={14186-14200},
  year={2018},
}

@article{lee2020automated,
  title={Automated detection of part quality during two\nobreakdash-photon lithography via deep learning},
  author={Lee, Xian Yeow and Saha, Sourabh K and Sarkar, Soumik and Giera, Brian},
  journal={Additive Manufacturing},
  volume={36},
  pages={101444},
  year={2020},
  publisher={Elsevier}
}

@inproceedings{mourka2020machine,
author = {Areti Mourka and Georgios D. Barmparis and Dimitra Ladika and Vasileia Melissinaki and David Gray and Maria Farsari},
title = {Machine Learning predicts printing parameters for multi\nobreakdash-photon polymerization three\nobreakdash-dimensional direct laser writing (3D-DLW) (Conference Presentation).},
organization = {Proceedings of SPIE 11271, Laser 3D Manufacturing VII. San Francisco: SPIE},
year = {2020},
}

@article{yang2022machine,
  title={Machine\nobreakdash-learning\nobreakdash-enabled geometric compliance improvement in two\nobreakdash-photon lithography without hardware modifications},
  author={Yang, Y. H. and Kelkar, Varun A. and Rajput, Hemangg S. and Coariti, Adriana C Salazar and Toussaint Jr, Kimani C and Shao, Chenhui},
  journal={Journal of Manufacturing Processes},
  volume={76},
  pages={841--849},
  year={2022},
  publisher={Elsevier}
}

@article{liu20183d,
  title={3D printing of bioinspired liquid superrepellent structures},
  author={Liu, X. J. and Gu, Hongcheng and Wang, Min and Du, Xin and Gao, Bingbing and Elbaz, Abdelrahman and Sun, Liangdong and Liao, Julong and Xiao, Pengfeng and Gu, Zhongze},
  journal={Advanced Materials},
  volume={30},
  number={22},
  pages={1800103},
  year={2018},
  publisher={Wiley Online Library}
}

@manual{NanoGuide,
    title        = {NanoGuide -- IP-S},
    year         = {2023},
    note         = {at \url{https://support.nanoscribe.com/hc/en-gb}},
    organization = {Nanoscribe GmbH \& Co. KG},
}

@inproceedings{AIR,
author = {Matthias Eifler and J{\"o}rg Seewig and Julian Hering and Georg von Freymann},
title = {Calibration of z-axis linearity for arbitrary optical topography measuring instruments},
series = {Proceedings of SPIE 9525, Optical Measurement Systems for Industrial Inspection IX. Munich:},
publisher = {SPIE},
year = {2015},
}

@article{CIN_r,
  title={Define and measure the dimensional accuracy of two\nobreakdash-photon laser lithography based on its instrument transfer function},
  author={Dai, G. L. and Hu, Xiukun and Hering, Julian and Eifler, Matthias and Seewig, J{\"o}rg and von Freymann, Georg},
  journal={Journal of Physics: Photonics},
  volume={3},
  pages={034002},
  year={2021},
  publisher={IOP Publishing Ltd}
}

@misc{AIR-ARS-ISO,
  title={Geometrical product specification (\text{GPS}) – Surface texture: Areal – Part 70: Material measures: ISO 25178-70:2014},
  author={\text{ISO}},
  journal={ISO 25178–70},
  volume={},
  pages={},
  year={2014},
}

@article{Eifler.2018,
 author = {Eifler, Matthias and Hering, Julian and von Freymann, Georg and Seewig, J{\"o}rg},
 year = {2018},
 title = {Calibration sample for arbitrary metrological characteristics of optical topography measuring instruments},
 pages = {16609--16623},
 volume = {26},
 number = {13},
 journal = {Optics express},
 doi = {10.1364/OE.26.016609}
}

@article{Survey-DataAugmentation,
 author = {Shorten, Connor and Khoshgoftaar, Taghi M.},
 year = {2019},
 title = {A survey on Image Data Augmentation for Deep Learning},
 pages = {60},
 volume = {6},
 number = {},
 journal = {Journal of Big Data}
}

@inproceedings{Eifler.DataProcessing,
author = {Matthias Eifler and Julian Hering and Andrej Keksel and Georg von Freymann and J{\"o}rg Seewig},
title = {Towards a continuous frequency band chirp material measure for surface topography measuring instrument calibration},
series = {Proceedings of SPIE 11782, Optical Measurement Systems for Industrial Inspection XII},
publisher = {SPIE},
year = {2021},
}

@article{Abbott1933,
 author = {Abbott, Ernest James and Firestone, Floyd A.},
 year = {1933},
 title = {Specifying surface quality - a method based on accurate measurement and comparison},
 pages = {569–572},
 volume = {55},
 journal = {Journal of Mechanical Engineering}
}

@manual{NanoFocus,
    title        = {NanoFocus -- µSurf Operating Manual},
    year         = {2003},
    month        = {June},
    organization = {NanoFocus AG},
    note         = {(Delivered with the device. Not available online.)}
}

@inproceedings{AlexNet,
 author = {Krizhevsky, Alex and Sutskever, Ilya and Hinton, Geoffrey E},
 title = {ImageNet Classification with Deep Convolutional Neural Networks}, 
 series = {Proceedings of the 26th International 
 Conference on Neural Information Processing Systems. Lake Tahoe:},
 publisher = {Curran Associates, Inc.,},
 year = {2012}
}

@article{VGG,
 author = {Simonyan, Karen and Zisserman, Andrew},
 year = {2015},
 title = {VERY DEEP CONVOLUTIONAL NETWORKS FOR LARGE\nobreakdash-SCALE IMAGE RECOGNITION},
 note = {Print at https://doi.org/10.48550/arXiv.1409.1556}
}

@article{Skliutas,
  title={Multiphoton 3D lithography},
  author={Skliutas, Edvinas and Merkininkaitė, Greta and Maruo, Shoji and Zhang, Wenxin and Chen, Wenyuan and Deng, Weiting and Greer, Julia and von Freymann, Georg and Malinauskas, Mangirdas},
  journal={Nature Reviews Methods Primers},
  volume={5},
  pages={15},
  year={2025},
}

@article{Torok:95,
author = {P. T\"{o}r\"{o}k and P. Varga and Z. Laczik and G. R. Booker},
journal = {Journal of the Optical Society of America A},
number = {2},
pages = {325--332},
publisher = {Optica Publishing Group},
title = {Electromagnetic diffraction of light focused through a planar interface between materials of mismatched refractive indices: an integral representation},
volume = {12},
year = {1995},
url = {https://opg.optica.org/josaa/abstract.cfm?URI=josaa-12-2-325},
doi = {10.1364/JOSAA.12.000325},
}

@article{Torok:95_errata,
author = {P. T\"{o}r\"{o}k and P. Varga and Z. Laczik and G. R. Booker},
journal = {Journal of the Optical Society of America A},
publisher = {Optica Publishing Group},
title = {Electromagnetic diffraction of light focused through a planar interface between materials of mismatched refractive indices: an integral representation: errata},
volume = {12},
number = {7},
pages = {1605--1605},
year = {1995},
url = {https://opg.optica.org/josaa/abstract.cfm?URI=josaa-12-7-1605},
doi = {10.1364/JOSAA.12.001605},
}

@article{polarization,
author = {Rek\^{s}tyt\.e, Sima and Jonavi\v{c}ius, Tomas and Gailevi\v{c}ius, Darius and Malinauskas, Mangirdas and Mizeikis, Vygantas and Gamaly, Eugene G. and Juodkazis, Saulius},
title = {Nanoscale Precision of 3D Polymerization via Polarization Control},
journal = {Advanced Optical Materials},
volume = {4},
number = {8},
pages = {1209-1214},
doi = {https://doi.org/10.1002/adom.201600155},
url = {https://advanced.onlinelibrary.wiley.com/doi/abs/10.1002/adom.201600155},
eprint = {https://advanced.onlinelibrary.wiley.com/doi/pdf/10.1002/adom.201600155},
year = {2016}
}

@article{pearson1895vii,
  title={VII. Note on regression and inheritance in the case of two parents},
  author={Pearson, Karl},
  journal={Proceedings of the royal society of London},
  volume={58},
  number={347-352},
  pages={240--242},
  year={1895},
  publisher={The Royal Society London}
}

@INPROCEEDINGS{OptimizerTest,
  author={Dogo, E. M. and Afolabi, O. J. and Nwulu, N. I. and Twala, B. and Aigbavboa, C. O.},
  title={A Comparative Analysis of Gradient Descent-Based Optimization Algorithms on Convolutional Neural Networks},
  series = {Proceedings of 2018 International Conference on Computational Techniques, Electronics and Mechanical Systems. Belgaum:}, 
  year={2018},
  publisher = {IEEE}
  }

@misc{Adam,
      title={Adam: A Method for Stochastic Optimization}, 
      author={Diederik P. Kingma and Jimmy Ba}, 
      note = {Print at https://doi.org/10.48550/arXiv.1412.6980},
      year={2017},

}

@misc{KerasAdam,
  title={\text{Keras - adam}},
  year={},
  howpublished={\url{https://keras.io/api/optimizers/adam/}},
}

@article{Nesterov,
  title={A method for solving a convex
 programming problem with convergence rate
 \text{O (1/$K^{2}$).}},
  author={Nesterov, Yurii E.},
  journal={Soviet Mathematics Doklady},
  volume={27},
  pages={372-376},
  year={1983},
}

@InProceedings{Nadam,
  author = 	 {Sutskever, Ilya and Martens, James and Dahl, George and Hinton, Geoffrey},
  title = 	 {On the importance of initialization and momentum in deep learning},
  series = {Proceedings of the 30th International Conference on Machine Learning. Atlanta:},
  publisher = {JMLR.org},
  year = 	 {2013},

}

@misc{KerasNadam,
  title={\text{Keras - Nadam}},
  year={},
  howpublished={At \url{https://keras.io/api/optimizers/Nadam/}},
}

@article{Cantoni,
author = {Cantoni, Federico and Maher, Daniel and Bosler, Eugenia and Kühne, Stefan and Barbe, Laurent and Oberschmidt, Dirk and Marquette, Christophe and Taboryski, Rafael and Tenje, Maria and Bunea, Ada-Ioana},
journal = {Additive Manufacturing},
title = {Round\nobreakdash-robin testing of commercial two\nobreakdash-photon polymerization 3D printers},
volume = {76},
pages = {103761},
year = {2023},
}

@Article{Ladika,
title = {X\nobreakdash-photon 3D lithography by fs\nobreakdash-oscillators: wavelength\nobreakdash-independent and photoinitiator\nobreakdash-free},
journal = {Light: Advanced Manufacturing},
volume = {5},
pages = {48},
year = {2024},
author = {Dimitra Ladika and Antanas Butkus and Vasileia Melissinaki and Edvinas Skliutas and Elmina Kabouraki and Saulius Juodkazis and Maria Farsari and Mangirdas Malinauskas},
}

@misc{Tensorflow,
title={ {TensorFlow}: Large\nobreakdash-Scale Machine Learning on Heterogeneous distributed Systems},
note={Print at https://doi.org/10.48550/arXiv.1603.04467},
author={
    Mart\'{i}n~Abadi and
    Ashish~Agarwal and
    Paul~Barham and
    Eugene~Brevdo and
    Zhifeng~Chen and
    Craig~Citro and
    Greg~S.~Corrado and
    Andy~Davis and
    Jeffrey~Dean and
    Matthieu~Devin and
    Sanjay~Ghemawat and
    Ian~Goodfellow and
    Andrew~Harp and
    Geoffrey~Irving and
    Michael~Isard and
    Yangqing Jia and
    Rafal~Jozefowicz and
    Lukasz~Kaiser and
    Manjunath~Kudlur and
    Josh~Levenberg and
    Dandelion~Man\'{e} and
    Rajat~Monga and
    Sherry~Moore and
    Derek~Murray and
    Chris~Olah and
    Mike~Schuster and
    Jonathon~Shlens and
    Benoit~Steiner and
    Ilya~Sutskever and
    Kunal~Talwar and
    Paul~Tucker and
    Vincent~Vanhoucke and
    Vijay~Vasudevan and
    Fernanda~Vi\'{e}gas and
    Oriol~Vinyals and
    Pete~Warden and
    Martin~Wattenberg and
    Martin~Wicke and
    Yuan~Yu and
    Xiaoqiang~Zheng},
  year={2016},
}

@misc{Keras,
  title={A superpower for ML developers},
  year={2015},
  howpublished={At \url{https://keras.io}},
}
\nocite{*}

\clearpage
\section{Appendix}

\subsection{MSE as Loss Function}\label{sec:LossFunction-MSE}
\begin{figure}[!ht]
    \centering
    \includegraphics[width=0.95\textwidth]{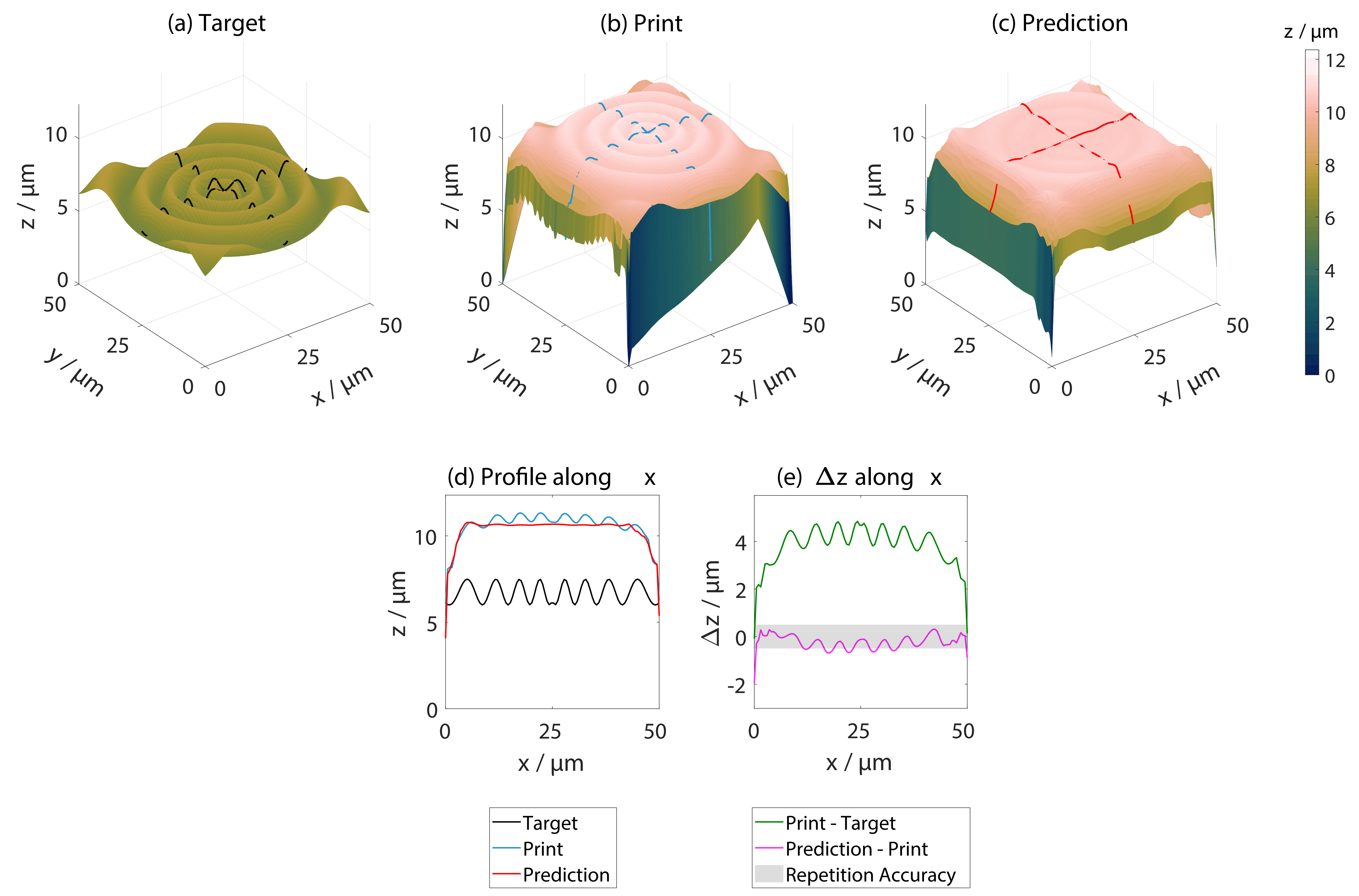}
    \caption{\textbf{Neural network prediction}. Comparison of an ideal radially symmetric $\text{CIN}_{\text{r}}$ type structure (a) with its corresponding 3D print (b) and the prediction of the neural network (c). The network was trained using mean square error as loss function. Profiles along the \textit{x}-axis and the deviations are shown in (d) and (e), respectively.}
    \label{fig:MSE-LossFunction}
\end{figure}

As mentioned earlier, the loss function is crucial for the training process of a neural network. Not only the convergence of the training but also the learned features and hence the network performance is strongly affected. Here, the widely used mean square error MSE as loss function is evaluated. While the MSE is a simple loss function and in many cases well suited, the complexity of our approach cannot be mapped. As exemplary visualized in Figure \ref{fig:MSE-LossFunction}, the MSE function is able to predict the average structure height. However, especially small features vanish due to averaging over the while structure surface. Therefore, a more complex loss function is needed. The illustration emphasizes the importance of the loss function and reaffirms our choice of the loss function as described in chapter \ref{sec:Lossfunction}.

\subsection{Training Data}\label{sec:Training Data}
\begin{table*}
    \centering
    \caption{\textbf{Training structures without DLW parameters.} 
    This dataset serves as basic training data without variation of DLW parameters. Slicing and hatching distances of 0.1\,µm as well as the laser power at roughly 45\,mW are kept constant. All 628 structures are physically fabricated and no data augmentation is used.}
    \label{tab:TrainStructuresNoParamVar}
    \begin{tabular}{rcll}
        \hline\hline
        \textbf{structure type} & \textbf{\# of samples} & \textbf{varied parameters} & \textbf{value range}\\
        \hline\hline
        block -- corrected block & 61 -- 48 & minimum height & (2 -- 17)\,µm \\
        \hline
        step -- corrected step & 70 -- 48 & \makecell[l]{minimum height \\ step height \\ step position} & \makecell[l]{(5 -- 10)\,µm \\ (0.5 -- 4)\,µm \\ \textit{x}\,=\,(16 -- 34)\,µm} \\
        \hline
        $\text{CIN}_{\text{r}}$-type -- corrected $\text{CIN}_{\text{r}}$-type (radial) & 227 -- 43 & \makecell[l]{minimum height \\ amplitude \\ minimum frequency \\ maximum frequency} & \makecell[l]{(2 -- 10.5)\,µm \\ (0.4 -- 1)\,µm \\ (0.005 -- 0.06)\,µm$^{-1}$ \\ (0.1 -- 0.4)\,µm$^{-1}$} \\
        \hline
        cosine structure ($x$-direction) & 23 & \makecell[l]{minimum height \\ amplitude \\ frequency} & \makecell[l]{(2 -- 8)\,µm \\ (0.5 -- 3)\,µm \\ (0.02 -- 0.075)\,µm$^{-1}$} \\
        \hline
        CIN ($x$-direction) & 15 & amplitude & (0.5 -- 0.85)\,µm \\
        \hline
        radially symmetric step & 20 & \makecell[l]{step height \\ step diameter} & \makecell[l]{(-4 -- 4)\,µm \\ (10 -- 20)\,µm} \\
        \hline
        array of small blocks on a large block & 13 & \makecell[l]{\# of blocks \\ height \\ edge length} & \makecell[l]{1\,x\,1 -- 3\,x\,3 \\ (3 -- 7)\,µm \\ (2 -- 20)\,µm} \\
        \hline
        various shapes on a large block & 30 & \makecell[l]{minimum height of large block \\ height of additional structures} & \makecell[l]{(2 -- 12)\,µm \\ (0.25 -- 3)\,µm}\\
        \hline
        AIR-type & 30 & \makecell[l]{minimum height \\ height scaling } & \makecell[l]{(2 -- 10)\,µm \\ (1 -- 5)\,µm}\\
        \hline\hline
    \end{tabular}
\end{table*}

\begin{table}
    \centering
    \caption{\textbf{Additional training data}.
    This data is implemented to cover for the DLW parameters, added as input or output of the neural networks. For all structures, the slicing and hatching distance was varied between (0.05 -- 0.7)\,µm and the laser power was varied between approximately (24 -- 60)\,mW. Here, in total 2765 structures are physically fabricated, including the 628 data sets shown in Tab. \ref{tab:TrainStructuresNoParamVar}. The remaining 2452 structures are generated using data augmentation.}
    \label{tab:ExtendedTrainingDatasets}
    \begin{tabular}{rc}
        \hline\hline
        \textbf{structure type} & \textbf{\# of samples} \\
        \hline\hline
        block -- corrected block & 219 -- 23 \\
        step -- corrected step & 224 -- 19 \\
        staircase structure & 167 \\
        CIN-type -- corrected CIN-type & 291 -- 10 \\
        AIR-type & 257 \\
        cosine structure ($x$-direction) & 290 \\
        radially symmetric step & 168 \\
        wedge-shaped structure & 168 \\
        paraboloid structure & 165 \\
        various shapes on a large block & 168 \\
        random structures & 28 \\
        data augmentation & 2452 \\
        \hline\hline
    \end{tabular}
\end{table}

\begin{figure}[b!]
    \centering\includegraphics[width=0.725\textwidth]{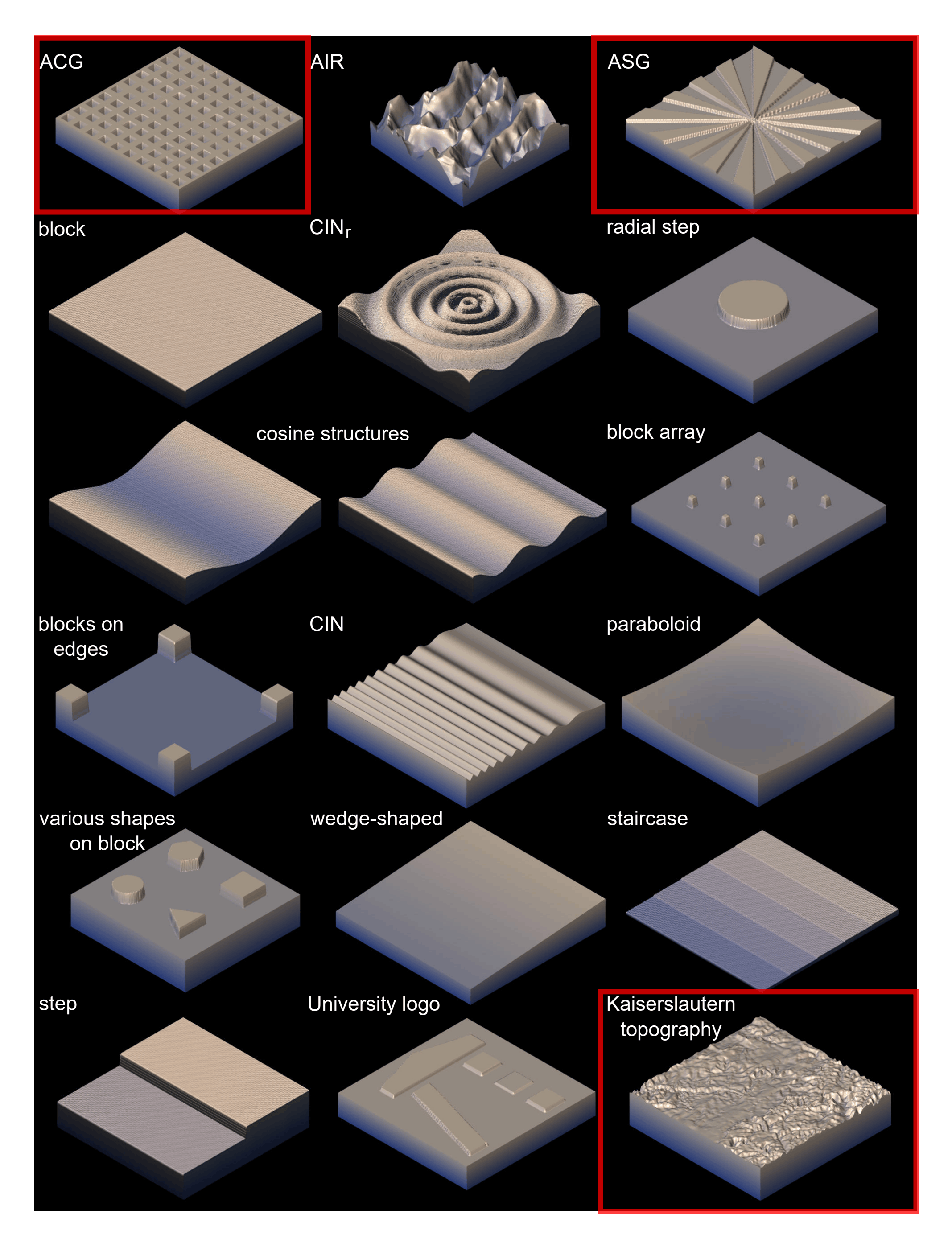}
    \caption{\textbf{CAD schemes.}
    Exemplary 3D illustrations of the structures, used for training and testing and developing of the neural networks. The ACG, ASG and topography structures marked in red were used to develop and test the models and the measurement and data processing workflow. Because of more frequent occurrence of errors due to lack of resolution or measurement artefacts, those structures were removed from training data. As measurement artefacts lead to a distorted analysis of the network performance, the results are not published as benchmark.}
    \label{fig:StructuresOverview}
\end{figure}

\noindent As listed in Tab.\,\ref{tab:TrainStructuresNoParamVar}, several different types of structures are used for training the neural networks.
Thereby, each structure has been generated with varying parameters, such as its height or the dimension of sub-features.
Hence, different amounts of each structure type have been included into the training datasets. In order to get an initial pool of training data, the 628 structures listed in Tab.\,\ref{tab:TrainStructuresNoParamVar} are fabricated. To keep the amount of varied parameters low for the first steps, DLW parameters are not varied in this case. Building on these 628 structures, the amount of training data is increased to implement the DLW parameters hatching \& slicing distance and laser power into the machine learning approaches. Therefore, 2137 additional structures are fabricated. While the structure types stay similar to the first 628, this time DLW parameters are varied. Hence, for similar structure types, the influence of the DLW parameters can be extracted from the new data sets. To further increase the amount of training data, 2452 more sets are generated using data augmentation. In total 5217 data sets are available, listed in Tab.\,\ref{tab:ExtendedTrainingDatasets}.\\
\noindent Some of those structures are exemplary illustrated as CAD images in Fig.\,\ref{fig:StructuresOverview}.

\subsection{Padding Influence}\label{sec:PaddingInfluence}
\begin{figure}[b!]
    \centering
    \includegraphics[width=0.90\textwidth]{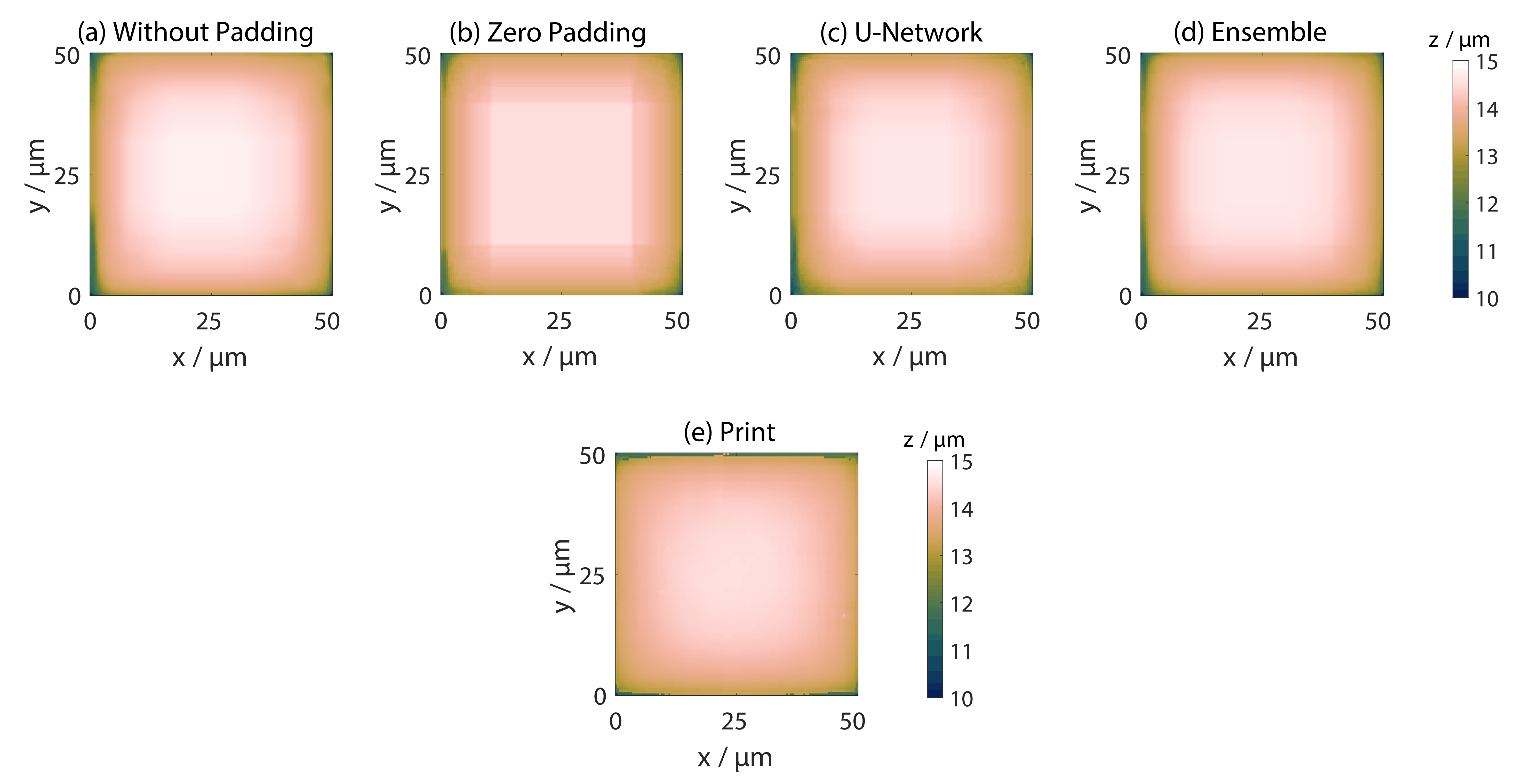}
    \caption{\textbf{Comparison of neural network architectures.}
    Prediction of a block by different architectures regarding padding and image sizes throughout the neural networks (a-c), where each architecture shows specific characteristics in its prediction result.
    Combined result of the four different approaches via ensemble learning is shown in (d). 
    The actual printing result is shown in (e).}
    \label{fig:Padding_Influence}
\end{figure}

\noindent As described in the main manuscript, padding  does have an influencing factor on the neural networks' outputs.
Figure\,\ref{fig:Padding_Influence} illustrates these influences onto the results of different neural network architectures and padding options in comparison to the actual measurement.

\subsection{Iterative Correction}\label{sec:Iterative Correction}
\begin{figure}[!ht]
    \centering
    \includegraphics[width=0.95\textwidth]{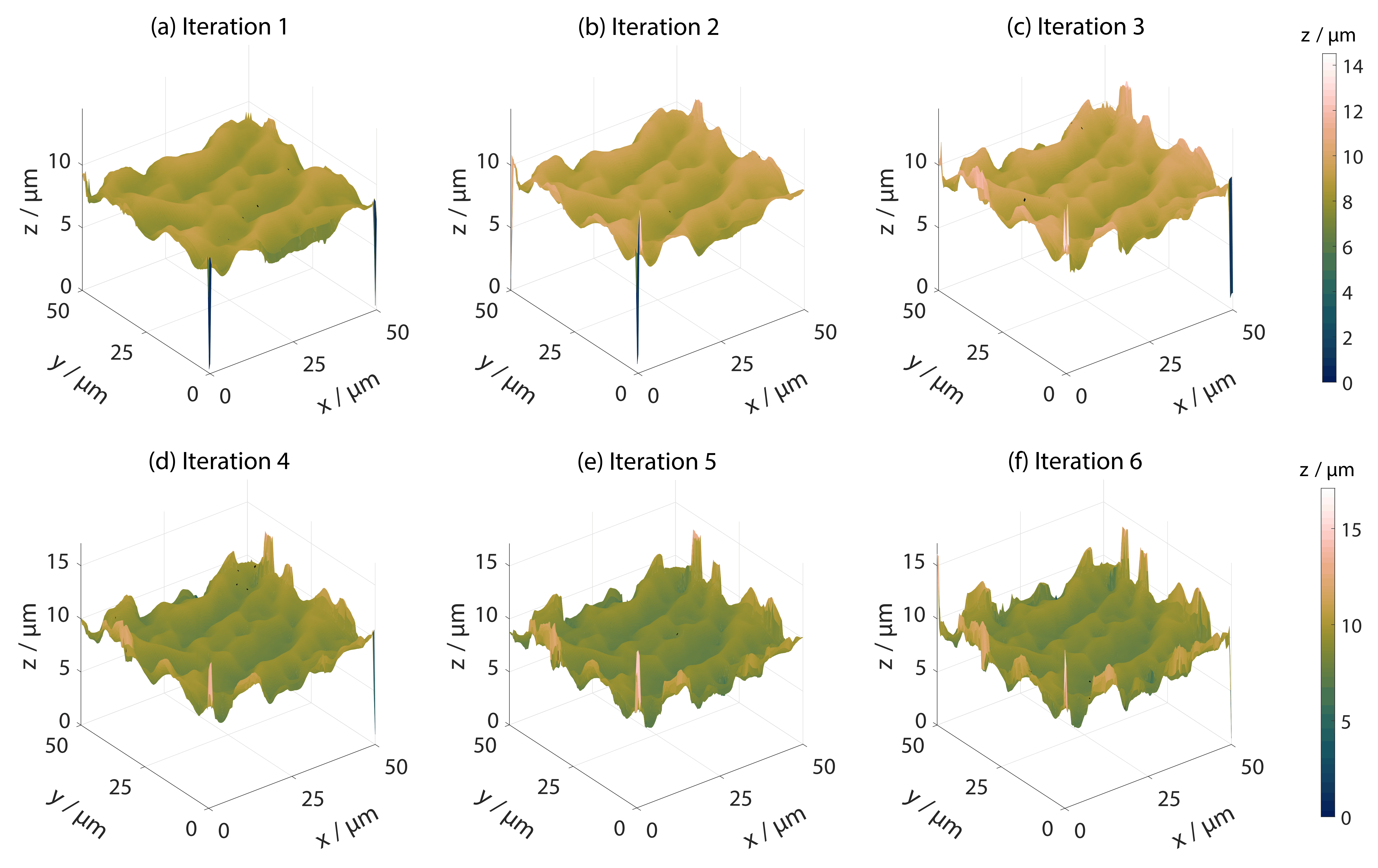}
    \caption{\textbf{Iterative Correction Approach.}
    Shown are the neural network's corrected printing results for the first six iterations of an AIR-type structure. 
    For higher iteration steps the increasing influence of artefacts, especially at the structure's edges becomes visible.}
    \label{fig:Iteration_Corrected_print_compressed}
\end{figure}

\noindent Since the \textbf{Iterative Correction} (\textbf{IC}) approach leads to enhancement of random artefacts, this is exemplary shown in Fig.\,\ref{fig:Iteration_Corrected_print_compressed} for the first six iteration steps.
The artefacts are most prominent at the edges of a structure.

\subsection{Computation}\label{sec:Computation}
\noindent The neural networks were implemented in Python using TensorFlow \cite{Tensorflow}, an open-source machine learning platform, in combination with the API Keras \cite{Keras}. 
For further data handling NumPy, SciPy, and scikit-learn packages were additionally used. 
In Tab.\,\ref{tab:Rechenressourcen+Dauer} the computation times to train a network are listed in dependence on the amount of datasets, training iterations and utilized hardware.
\begin{table}[!b]
    \centering
    \caption{\textbf{Computation resources.}
    Overview over the resources used for training the different amount of datasets.}
    \label{tab:Rechenressourcen+Dauer}
    \begin{adjustbox}{width=\textwidth}
    \begin{tabular}{cccccc}
        \hline\hline
        \textbf{\# of training datasets} & \textbf{iterations} & \textbf{used RAM / GB} & \textbf{used CPUs} & \textbf{used GPUs} & \textbf{training duration / d:h:m} \\
        \hline\hline
        497 & 1000 & 1.5 & 8 & - & 0:11:59 \\
        568 & 1000 & 1.6 & 8 & - & 0:14:41 \\
        628 & 1000 & 1.8 & 4 & - & 1:14:15 \\
        1886 & 500 & 6.1 & 8 & 1x Volta V100 & 1:14:44 \\
        2779 & 500 & 8.8 & 12 & 1x Volta V100 & 2:00:42 \\
        5217 & 1000 & 18.5 & 16 & 1x Volta V100 & 6:13:44 \\
        \hline\hline
    \end{tabular}
    \end{adjustbox}
\end{table}

\begin{figure}[t]
    \centering\includegraphics[width=0.95\textwidth]{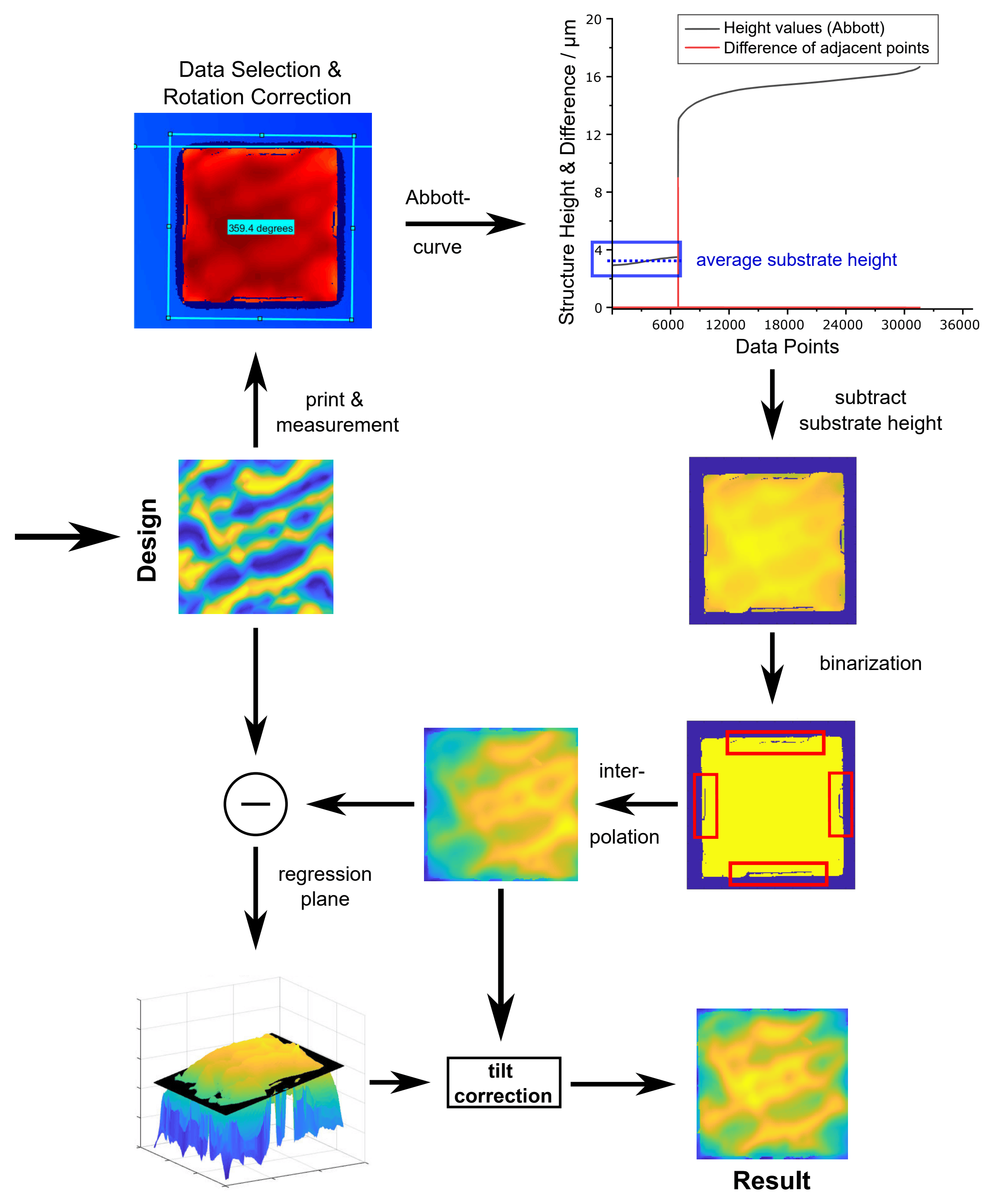}
    \caption{\textbf{Data Processing Scheme.}
    Schematic visualization of the data processing steps as described in chapter \ref{subsec:dataprep} using the example of an AIR structure.}
    \label{fig:DataProcessingScheme}
\end{figure}

\end{document}